\documentclass[letterpaper]{JHEP3}
\usepackage{amsmath,xspace}
\usepackage[pdftex]{graphicx}
\pdfoutput=1

\newcommand{\jb}{{\bar\jmath}}

\newcommand{\R}{\mathbb{R}}
\newcommand{\C}{\mathbb{C}}
\newcommand{\Z}{\mathbb{Z}}
\newcommand{\CP}{\mathbb{CP}}
\newcommand{\CPN}{$\mathbb{CP}^N$\xspace}

\newcommand{\K}{K\"ahler\xspace}

\title{Energy functionals for Calabi-Yau metrics}

\author{
Matthew Headrick and Ali Nassar \\ 
Martin Fisher School of Physics, Brandeis University, Waltham MA 02454, USA \\ 
\email{mph@brandeis.edu}, \email{anassar@brandeis.edu}
}

\abstract{
We identify a set of ``energy" functionals on the space of metrics in a given \K class on a Calabi-Yau manifold, which are bounded below and minimized uniquely on the Ricci-flat metric in that class. Using these functionals, we recast the problem of numerically solving the Einstein equation as an optimization problem. We apply this strategy, using the ``algebraic" metrics (metrics for which the \K potential is given in terms of a polynomial in the projective coordinates), to the Fermat quartic and to a one-parameter family of quintics that includes the Fermat and conifold quintics. We show that this method yields approximations to the Ricci-flat metric that are exponentially accurate in the degree of the polynomial (except at the conifold point, where the convergence is polynomial), and therefore orders of magnitude more accurate than the balanced metrics, previously studied as approximations to the Ricci-flat metric. The method is relatively fast and easy to implement. On the theoretical side, we also show that the functionals can be used to give a heuristic proof of Yau's theorem.
}

\preprint{BRX-TH-612}

\begin{document}

\section{Introduction}

Calabi-Yau (CY) manifolds play a pivotal role in complex differential geometry, algebraic geometry, and string theory. Part of the reason for their importance, for both mathematics and physics, lies in the Ricci-flat metrics they admit. The existence of these metrics is guaranteed by Yau's theorem; this theorem, however, is not constructive, and, outside of flat metrics and orbifolds thereof, no exact solutions to the Einstein equation are known on any CY manifold. This state of affairs naturally leads to the question of whether it is feasible to compute useful numerical approximations to the Ricci-flat metrics, a question that has given rise to a small literature in recent years \cite{Headrick:2005ch,MR2508897,Douglas:2006hz,Douglas:2006rr,Braun:2007sn,Braun:2008jp,KellerLukic}. A few different methods have been put forward, all of which make crucial use of the advantages offered by \K geometry compared to real geometry.\footnote{Related methods for solving the Einstein equation on toric manifolds were discussed in \cite{Doran:2007zn, MR2540879, MR2483361}.} The purpose of this paper is to introduce a new method for this problem, that scores well on all three of the major considerations that make such a method useful: ease of implementation, scale of computational resources required, and accuracy of the approximations obtained. The method is based on a mathematical observation that may be of some interest in its own right.

The mathematical and physical motivations for solving the Einstein equation on CYs are discussed at length in the papers cited above, and we will not repeat them here. However, it is important to note that the scope of the potential applications depends strongly on the performance of the methods available. For example, for some problems it is sufficient to compute the metric at a single point in the CY moduli space, while for others it is necessary to scan over a part of the moduli space, requiring much faster methods.

Any numerical method for solving a PDE must address two issues. The first is how to represent, or store, the function being solved for---in this case, the \K potential---for example using a real-space lattice or a spectral representation. The second is, given that representation, how to solve the equation, or, more precisely, how to find the best approximation within that representation to the exact solution. For the problem at hand, the principal challenges in terms of the representation are the high dimensionality of the manifold (usually four or six real dimensions, so that even for moderate resolutions storage quickly becomes a limiting factor) and its typically very complicated topology. The principal challenge in terms of solving the Einstein equation is its nonlinearity.

The first methods developed for solving the Einstein equation on CYs, by Headrick and Wiseman \cite{Headrick:2005ch}, used a real-space lattice representation of the \K potential, and solved the equation by a Gauss-Seidel relaxation method. The method was straightforward, if not particularly elegant. One downside was the messiness of dealing with coordinate patches. The main factor limiting the accuracy of the metrics obtained was storage, due to the large number of lattice points required for high resolution in high dimensions (the method was tested on K3; storage would have severely limited the possible resolutions for threefolds).

Subsequently, Donaldson proposed a radically different set of methods based on a spectral (momentum-space) representation of the metric \cite{MR2508897}. Specifically, he advocated the use of so-called algebraic metrics, which had been studied from a theoretical viewpoint by Tian and others. We will review the definition, motivation, and properties of algebraic metrics in Section 3. The salient properties for our present purposes are (1) they are based on polynomials, and therefore very easy to work with both theoretically and computationally; (2) they eliminate the messiness of coordinate patches; and (3) within a certain \K class they are capable of approximating any given smooth metric---including the Ricci-flat one---exponentially well in $k$, the degree of the polynomials ($k$ is analogous to the highest mode number of a finite set of Fourier modes used to approximate a function, for example, on $S^1$). The latter property essentially promises a form of data compression, and, when applied to the Ricci-flat metric, has potential to solve the storage problem mentioned above.

As with any spectral representation, the challenge then comes in solving the PDE. Linear PDEs (especially homogeneous ones) become simpler when written in momentum space, as the different modes decouple from each other. For nonlinear ones such as the Einstein equation, on the other hand, every mode is coupled to every other mode, leading to a complicated system of nonlinear equations. Donaldson proposed two kinds of approximate solutions to the Einstein equation within the framework of the algebraic metrics. The first was the so-called balanced metrics. For each $k$ the balanced metric is the unique solution to a certain integral equation, which can be found by iterating a certain integral map.\footnote{The balanced metrics are of significant mathematical and possibly physical interest quite apart from the problem of solving the Einstein equation \cite{MR2508897,Douglas:2008es}.} Donaldson showed that they approach the Ricci-flat metric as $k\to\infty$, but only as a power of $k$. Thus they do not achieve the exponential accuracy promised by the algebraic metrics. The second proposal, called ``refined" metrics, were based on a Galerkin method: the error in solving the Einstein equation (more precisely, the equivalent Monge-Amp\`ere equation) was required to be orthogonal to the full set of basis functions (in this case, the degree-$k$ polynomials). The refined metrics are expected to convergence exponentially to the Ricci-flat metric. However, efficiently computing the refined metric proved difficult.

%These are algebraic metrics that satisfy a certain integral equation, and were shown by Donaldson to approach the Ricci-flat metric as $k\to\infty$, although only as a power of $k$; thus they do not achieve the exponential accuracy possible with the algebraic metrics \cite{MR2508897}. The balanced metrics (which are of significant mathematical and possibly physical interest quite apart from the problem of solving the Einstein equation \cite{MR2508897,Douglas:2008es}) can be computed by iterating a certain integral map on the matrix $h$ \cite{MR2508897}. 

Now, it is well known in numerical analysis that it is far easier to minimize a well-behaved function than to solve a non-linear set of equations.\footnote{For a standard discussion of this point, see e.g.\ section 9.6.1 of \cite{nr}.}  Here, by ``well-behaved", we mean smooth, bounded below, and having no critical points other than a single global minimum.This is true both from a theoretical viewpoint and in terms of actual available algorithms; while there exist several efficient and robust numerical optimization algorithms, for solving non-linear equations there is essentially only the Newton-Raphson method, which has poor global convergence properties, and a few variants on it. So it is nearly always advantageous, when possible, to convert the latter type of problem into the former. In the case of an elliptic PDE, a well-behaved functional whose variational derivative yields the PDE is usually called an energy functional. 
%In working with an elliptic PDE, either theoretically or computationally, a very useful strategy is to formulate it as a variational problem (in numerical jargon, an optimization problem), by finding an ``energy" functional that is bounded below and globally minimized on the solution to the PDE; better if the energy functional has no local minima, and best if it has no other critical points at all. We will refer to such a functional as \emph{well-behaved}. 
Note that most elliptic PDEs do not admit such a variational formulation. (The utility of such a formulation is also of course clear from the mathematical point of view, for example for proving the existence of solutions.) Given an energy functional and a representation scheme (real-space, spectral, or otherwise), one minimizes the energy functional within the corresponding finite-dimensional space of functions.\footnote{Note that the error in solving the equation will necessarily be orthogonal, in the full function space, to the subspace in which one is working. Hence any method of this type automatically solves some Galerkin-type condition.} The prototype for this method is the Rayleigh-Ritz variational method for finding the smallest eigenvalue of a linear operator. Especially for spectral representations, this strategy can vastly simplify the problem of solving the PDE.

In order to follow this strategy for case of the Einstein equation on a CY manifold,\footnote{The possibility of following a functional-minimization strategy was briefly mentioned in \cite{Douglas:2006rr}.} in Section 2 we define a family of energy functionals on the space of \K metrics on a given CY manifold, that are bounded below and minimized within each \K class precisely on the Ricci-flat metric.\footnote{These functionals share some of the properties of  the Calabi energy \cite{MR645743}, the Mabuchi K-energy \cite{MR867064}, and its generalizations \cite{MR2540879,Berman}, although they are somewhat simpler.} The existence of these functionals depends crucially on the magic of \K geometry and should by no means be taken for granted---comparable functionals do not, as far as we know, exist in the Riemannian setting (for example, the Einstein-Hilbert action is unbounded below, while the integrated square of the Ricci tensor has many spurious local minima). Note that the functionals we define are not tied to the algebraic metrics, and could be used to solve the Einstein equation given any spectral or other representation scheme. We show, incidentally, that these functionals can also be used to give a heuristic proof of Yau's theorem. The argument has loopholes, as we discuss, but may nonetheless be useful as a quick way to understand why one should expect Yau's theorem to hold.\footnote{The recent paper \cite{Berman} uses a variational approach based on the Mabuchi K-energy \cite{MR867064} to prove a weak version of Yau's theorem.}

In Section 4 we combine the algebraic metrics with one of our energy functionals to define ``optimal" metrics. We derive the Galerkin condition for the optimal metrics, and compare it to the one defining Donaldson's refined metrics.

In Section 5 we report the results of applying our method to the Fermat quartic and to a one-real-parameter family of quintics that includes the Fermat and conifold quintics. The method is relatively simple to implement (requiring just a few pages of \emph{Mathematica} code), and very efficient (running for a few minutes on a  laptop computer, it yields metrics that are likely accurate enough for most applications). In particular, the method achieves the exponential accuracy promised by the algebraic metrics. As a result, the metrics found are far more accurate, for a given value of $k$, than the balanced metrics. (The high-accuracy metrics we obtain on the compact conifold and deformed conifold, in particular, are perhaps relevant for certain models of string phenomenology; see \cite{Baumann:2010sx} and references therein). Some aspects of our implementation of the method are described in Section 6.

\subsection{Future directions}

The work described in this paper can be extended in many different directions. The most obvious is to apply the same method to CYs with less symmetry. Doing so should not require significant changes to the method, although attaining high-accuracy metrics will probably call for the use of a compiled language rather than \emph{Mathematica}. As we discuss theoretically in subsection 3.2 and show in practice in subsection 5.2, for CYs on which the Ricci-flat metric contains small, highly-curved regions (e.g.\ near an orbifold or conifold point in the CY moduli space), the algebraic metrics may not provide the ideal representation of the metric. A real-space representation may be preferable. Nonetheless, the energy-minimization strategy described here should still apply.

A second direction would be to evaluate interesting geometric quantities given the metrics described here, such as curvature invariants and eigenvalues of various operators,\footnote{The papers \cite{Headrick:2005ch,Braun:2008jp} discuss the calculation of scalar Laplacian eigenvalues on numerical CY metrics, with the second reference specializing in particular to algebraic metrics.} and to compute the metric on the moduli space using the metric on the CY.\footnote{The recent paper \cite{KellerLukic} studied the calculation of the moduli space metric using the balanced metrics.}

A third direction would be to consider the Einstein equation on non-CY \K manifolds. This could in principle be doable using the Calabi energy \cite{MR645743} or the Mabuchi K-energy \cite{MR867064}.

A more difficult problem would be to incorporate matter fields and/or the higher-deriva\-tive corrections to the Einstein equation that occur in string theory (as for example was done for the hermitian Yang-Mills equation in \cite{Douglas:2006hz}). Here the main issue is whether the problem can still be expressed as the minimization of a well-behaved energy functional. Although we have not investigated this question, it seems likely that the answer will be yes as long as the geometry remains \K (and perhaps more generally in the context of ``generalized geometry").

Perhaps in combination with some of the above generalizations, one could employ the metrics derived by our method in models of string phenomenology, as was done in \cite{Douglas:2006hz,Braun:2007sn}.

Finally, although in this paper we mainly focus on the energy functionals as calculational tools, it is worth exploring whether they have any direct physical interpretation, either from a spacetime or a world-sheet (i.e.\ supersymmetric sigma model) viewpoint.

\section{Energy functionals}

Let $X$ be a compact $n$-dimensional \K manifold equipped with a fixed complex structure and a fixed volume form $\hat\mu$, which we take to be normalized, $\int_X\hat\mu=1$. Associated to each \K form $J$ is a volume form
\begin{equation}
\mu_J = \frac{J^n}{n!}\,.
\end{equation}
The ratio of the two volume forms is a positive scalar that, following Donaldson \cite{MR2508897}, we denote $\eta$:
\begin{equation}
\eta \equiv \frac{\mu_J}{\hat\mu}\,.
\end{equation}
Yau's theorem states that each \K class contains a (unique) representative on which $\eta$ is constant. In the following we will assume that the \K class is normalized, $\int_X\mu_J = 1$, so the distinguished representative has
\begin{equation}
\eta = 1\,.
\end{equation}
This is a second-order elliptic PDE of 
-Amp\`ere type for the \K potential.

In Sections 3--6, we will assume that $X$ is Calabi-Yau, and we will choose $\hat\mu$ to be the volume form associated with its holomorphic $(n,0)$-form $\Omega$:
\begin{equation}
\hat\mu = (-i)^n\Omega\wedge\bar\Omega\,.
\end{equation}
The Ricci tensor for $J$ is then $R_{i\jb} = -\partial_i\partial_\jb\ln\eta$, and the Einstein equation is equivalent to the Monge-Amp\`ere equation.

Using the metric alone, one can write down several natural and useful geometric functionals, such as the Einstein-Hilbert action $\int_X\mu_JR$ and the Calabi energy $\int_X\mu_JR^2$ \cite{MR645743}. (Note however that the Einstein-Hilbert action vanishes for any \K metric on a CY manifold.) The introduction of a background volume form, however, can be quite useful for constructing interesting geometric functionals, as shown by the example of Perelman's first functional \cite{perelman-2002} (discussed in subsection 2.2).\footnote{In the \K context one also has the Mabuchi K-energy  \cite{MR867064}, and various generalizations of it, as discussed for example in \cite{MR2540879,Berman}.} In subsection 2.1, we will see that, using such a volume form, even functionals with \emph{no} derivatives acting on the metric can have non-trivial properties. Then in subsection 2.2 we will consider a two-derivative functional similar to Perelman's. In Section 5 we will describe the results of numerical evaluation and minimization of both types of functional.

\subsection{Ultralocal functionals}

In this subsection we will show that a large class of functionals with no derivatives acting on the metric can yield the Monge-Amp\'ere equation (and therefore, in the CY case, the Einstein equation) as their Euler-Lagrange equation.

Let $F:\R^+\to\R$ be a differentiable convex function that is bounded below, and define
\begin{equation}
E_F[J] = \int_X \hat\mu\,F(\eta)\,.
\end{equation}
Clearly $E_F$ is bounded below as well. Adding a linear term to $F$ changes $E_F$ only by a constant ($E_{F+\alpha\eta}[J]=E_F[J]+\alpha$), so without loss of generality we may assume that $F(\eta)$ attains its minimum at $\eta=1$. It follows that, in each \K class, $E_F$ has a unique global minimum on the solution to the Monge-Amp\`ere equation.

We will now show that $E_F$ has no local minima or other critical points. Under variations of $J$ within its cohomology class,
\begin{equation}
\delta J = i\partial\bar\partial\phi\,,
\end{equation}
the change in $\eta$ to first order is
\begin{equation}\label{etavary}
\delta\eta = \frac12\eta\nabla_J^2\phi\,.
\end{equation}
Hence the variation of $E_F$ is\footnote{It is interesting to note that if we choose $F(\eta)=\eta\ln\eta$, so that $E_F$ is (minus) the von Neumann entropy of the probability distribution $\eta$ on $X$, then its gradient flow is the Calabi flow.}
\begin{equation}\label{varEF}
\delta E_F = \int_X\hat\mu\,F'(\eta)\delta\eta = 
\frac12\int_X\mu_JF'(\eta)\nabla^2_J\phi = 
\frac12\int_X\mu_J\nabla^2_JF'(\eta)\,\phi\,.
\end{equation}
The Euler-Lagrange equation is thus
\begin{equation}
\nabla^2_JF'(\eta) = 0\,,
\end{equation}
which, given the compactness of $X$ and the positivity of the metric, is equivalent to the constancy on $X$ of $F'(\eta)$, which in turn implies the Monge-Amp\`ere equation.

There is another way to think about this, which shows that the assumption of differentiability of $F$ is superfluous. Yau's theorem establishes a one-to-one mapping between a given normalized \K class and the space of positive functions $\eta$ on $X$ obeying $\int_X\hat\mu\,\eta = 1$. Given a convex function $F(\eta)$ that is minimized at $\eta=1$, it is clear that the global minimum (and only critical point) of $\int_X\hat\mu\,F(\eta)$ on that function space---and therefore on the \K class---is $\eta=1$.

Variational formulations of elliptic PDEs can often be used to establish the existence of solutions. Therefore, rather than \emph{assuming} Yau's theorem, it is interesting to ask whether we can use such a functional to \emph{prove} it.\footnote{The recent paper \cite{Berman} uses a variational approach based on the Mabuchi K-energy \cite{MR867064} to prove a weak version of Yau's theorem.} Of course, a rigorous proof would be far outside the scope of this paper, but we would like to point out a simple heuristic argument, or ``physicist's proof", using the fact that $E_F$ is extremized precisely on solutions. The general principle involved is that a differentiable real function defined on an open set, that is bounded below and goes to infinity on the boundary of its domain, must have a minimum and therefore an extremum. In our case, the open set in question is the space of (smooth) \K metrics on $X$ in a given class. The non-trivial issue is thus the behavior of $E_F$ on the boundary of this space. The generic degeneration of the metric is the development of an isolated conical singularity, with $\eta$ going to 0 or $\infty$ as a power of $|z|^2$ in some local coordinate system. In order to guarantee the divergence of $E_F$ on such a singular metric, it is sufficient for $F(\eta)$ to go to infinity like (say) $e^\eta$ as $\eta\to\infty$, and $e^{1/\eta}$ as $\eta\to0$. Of course, there are also other (non-generic) kinds of degeneration, in which $\eta$ goes to 0 or $\infty$ sufficiently slowly that such an $E_F$ remains finite; indeed, for any choice of $F$, one can construct such a singular metric. Another possible degeneration is where some eigenvalues of the metric go to infinity and others go to zero simultaneously in such a way that $\eta$ remains finite; then of course $E_F$ will remain finite for any choice of $F$.

\subsection{A two-derivative functional}

Here we give an example of a functional that, like the Einstein-Hilbert action, contains two derivatives acting on the metric, but, unlike it, is bounded below:
\begin{equation}\label{Ehatdef}
\hat E[J] = 
\int_X\hat\mu\,g^{\jb i}\partial_i\ln\eta\,\partial_\jb\ln\eta\,.
\end{equation}
A short computation shows that
\begin{equation}\label{Ehatvar}
\delta\hat E = 4\int_X\hat\mu\,\eta^{1/2}\phi^{\jb i}\partial_i\partial_\jb (\eta^{-1/2})\,,
\end{equation}
where $\phi_{i\jb} \equiv \partial_i\partial_\jb\phi = \delta g_{i\jb}$. Hence $\hat E$ is stationary if and only if $\eta$ is constant.

In the Calabi-Yau case, with $\hat\mu = (-i)^n\Omega\wedge\bar\Omega$, it follows immediately that
\begin{equation}\label{Ehatiden}
\hat E = -\frac12\int_X\hat\mu\,R\,.
\end{equation}
This identity reveals a connection to Perelman's first functional \cite{perelman-2002}, which is defined for a Riemannian metric but, like $\hat E$, depends on a background volume form; in our present notation it would be written
\begin{equation}
E_P = 
-\int_X\hat\mu
\left(R + g^{ab}\partial_a\ln\eta\,\partial_b\ln\eta\right).
\end{equation}
Its gradient flow is Ricci flow supplemented by a diffeomorphism
\begin{equation}\label{Perelmanflow}
\dot g_{ab} = -R_{ab} - \nabla_a\partial_b\ln\eta\,.
\end{equation}
In the CY case, however, due to the identity \eqref{Ehatiden}, $E_P$ vanishes identically; correspondingly, the gradient flow \eqref{Perelmanflow} ($\dot g_{i\jb} = 0$, $\dot g_{ij} = -\nabla_i\partial_j\ln\eta$) is orthogonal to the space of \K metrics. The nice properties of $\hat E$ suggest that it may be natural to consider its gradient flow,
\begin{equation}
\dot g_{i\jb} = -\partial_i\partial_\jb (\eta^{-1/2})
\end{equation}
(compare this to Ricci flow, $\dot g_{i\jb} = \partial_i\partial_\jb\ln\eta$).

\section{Algebraic metrics}

In this section we will briefly review the definition of and motivation for the algebraic metrics. The discussion follows that in \cite{MR2508897}, but is hopefully translated into a more physicist-friendly language. More mathematically sophisticated discussions can be found in \cite{MR2508897,Douglas:2006rr}.

Essentially we are searching for a nice basis of functions on our Calabi-Yau $X$, in which to expand the \K potential. One's first thought might be to use, for example, the eigenfunctions of some Laplacian operator. But of course we don't know those eigenfunctions, which in any case probably do not have any particularly simple form. Instead the strategy is to embed $X$ in \CPN for some $N$, since \CPN is a nice space that admits a highly symmetrical canonical metric, namely the Fubini-Study (FS) metric; as we will see, the Laplacian eigenfunctions with respect to the FS metric take on a very simple form. Remarkably, these eigenfunctions remain simple even when pulled back to $X$, and provide an ideal basis in which to expand the \K potential.

We thus require $X$ to be embeddable as an algebraic variety in \CPN; furthermore, the \K class we consider on $X$ must be the one induced from  \CPN (hence it must be an element of the Picard lattice).\footnote{The method generalizes straightforwardly to algebraic varieties in products of projective spaces, weighted projective spaces, etc.} This is obviously a limitation; on the other hand, it includes many CYs of interest, both mathematically and physically.

\subsection{Algebraic metrics on \CPN}

Let the homogeneous coordinates on \CPN be $z^a$. Recall that the FS metric on \CPN has a \K potential given, in patch $b$ (i.e.\ where $z^b\neq0$), by
\begin{equation}\label{FS}
K_{(b)}^{\rm FS} = \ln\frac{\sum_a|z^a|^2}{|z^b|^2}\,.
\end{equation}
The denominator inside the logarithm ensures that the \K potential transforms correctly on the patch overlaps. The algebraic metrics are a straightforward generalization of the FS metric, in which we replace the numerator with a higher-degree homogeneous polynomial. Let $k$ be a positive integer, and $Z^A$ be the degree-$k$ monomials in the $z^a$; there are $N_k\equiv{N+k\choose k}$ of them. An algebraic metric is one whose \K potential is
\begin{equation}\label{amdef}
K_{(b)}^h = 
\ln\frac{\left(Z^Ah_{A\bar B}\bar Z^{\bar B}\right)^{1/k}}{\left|z^b\right|^2}
\end{equation}
for some positive-definite hermitian matrix $h$. It's easy to see that the \K class is independent of the choice of $k$ and $h$.\footnote{Note that this class is not volume-normalized (does not satisfy $\int\mu_J=1$), either on \CPN or when pulled back onto $X$.\label{normalization}} The matrix elements of $h$ are the degrees of freedom of the algebraic metric (except that multiplying $h$ by a positive number just shifts the \K potential by a constant, leaving the associated metric unchanged). Hence the larger $k$, the larger the space of algebraic metrics.

Another way to think about the algebraic metrics is in terms of the Veronese embedding of  \CPN into $\CP^{N_k}$, defined by the mapping $z\mapsto Z$. From this point of view $\ln(Z^Ah_{A\bar B}\bar Z^{\bar B}/|Z^C|^2)$ defines a generalized FS metric on $\CP^{N_k}$, and $K^h$ is ($1/k$ times) its pull-back to \CPN.

The algebraic metrics are a spectral representation in the following sense. It turns out that the $N_k^2$ functions $Z^A\bar Z^{\bar B}/(\sum_a|z^a|^2)^k$ are a basis for the first $k$ eigenspaces of the Laplacian (with respect to the FS metric) on \CPN, in analogy to the fact that the spherical harmonics on $S^2$ can be written as polynomials in the coordinates on an embedding $\R^3$ \cite{MR2508897}. Therefore in the limit $k\to\infty$ they form a complete basis for the functions on \CPN. Any metric in the same class as the FS metric has \K potential $K_{(b)} = K^{\rm FS}_{(b)} + \phi$ for some globally defined real function $\phi$. Hence when we write \eqref{amdef}, we are expanding $\phi$ (more precisely, $e^{k\phi}$) in a truncated basis of spherical harmonics, with the matrix elements of $h$ being the coefficients. Note that at large $k$ the number of degrees of freedom is $N_k^2\sim k^{2N}$, corresponding roughly to $k$ ``Fourier modes" in each real direction. As usual in Fourier analysis, if $\phi$ is smooth then the truncated sum of spherical harmonics will approximate it exponentially well in $k$ (the technical statement is that this sum will approximate it better than $k^\nu$ for any $\nu$  \cite{MR2508897}; in practice this usually means exponential convergence).

\subsection{Algebraic metrics on $X$}

Now let us return to $X$, which is sitting inside \CPN along the vanishing locus of one or more homogeneous polynomials in the $z^a$. The ``spherical harmonics" $Z^A\bar Z^{\bar B}/(\sum_a|z^a|^2)^k$ are linearly independent on \CPN, but not when restriced to $X$. However, it is very easy to see precisely which functions are redundant: the linear combinations of the $Z^A$ that vanish on $X$ are simply the polynomials that are proportional to (any of) the defining polynomial(s) of $X$. For example, in the case $N=n+1$, $X$ is defined by a single degree $N+1$ polynomial. If $k>N$, the space of degree $k$ polynomials proportional to that polynomial is $N_{k-(N+1)}$ dimensional. Hence the dimensionality of the quotient space is
\begin{equation}\label{Nkprime}
N'_k =
\left.\begin{cases} N_k, & k\le N \\ N_k - N_{k-(N+1)}, & k>N \end{cases}\right\} =
\left.\begin{cases} 2, & N=1 \\ 3k, & N=2 \\ 2(k^2+1), & N=3 \\ \frac56k(k^2+5), & N=4 \end{cases}\right\}
\sim k^n.
\end{equation}
Note that the (real) dimensionality of the space of algebraic metrics has gone down from $k^{2N}$ on \CPN to $k^{2n}$ on $X$.

The elegance and simplicity of the algebraic metrics leads immediately to  two major practical advantages. First, one does not have to deal with coordinate patches on $X$, only those on \CPN, which are quite simple. Second, they are based on polynomials, which are the easiest and fastest kind of function to evaluate numerically.

A final note concerning the expected accuracy of the algebraic metric: The estimate of exponential accuracy mentioned above concerns the \emph{asymptotic} accuracy for large $k$. One expects, however, that in order to get a good approximation, $k$ must be sufficiently large that the shortest wavelength of the Fourier modes involved (roughly $1/k$) is no larger than the smallest characteristic size of the function being approximated. Hence we expect the algebraic metrics to be most useful for CYs that are not just smooth but also geometrically uniform, i.e.\ that do not have large ratios of characteristic scales. We expect this to be the case far from singular points in the CY moduli space. CYs that are close to orbifold or conifold singularities in the moduli space, for example, will contain small, highly curved regions. In order to accomodate the multiple length scales occurring in such cases, it may be more suitable to use a representation scheme whose basis functions are localized, such as wavelets or finite elements.

\section{Optimal metrics}

The energy functionals described in Section 2 and the algebraic metrics described in Section 3 may be combined to yield a numerical method for  solving the Einstein equation on an algebraic Calabi-Yau manifold $X$: for a given $k$, minimize one of the functionals $E_F$ or $\hat E$ over the space of matrices $h$ appearing in the algebraic metric, using an appropriate optimization algorithm.

%In this section we will describe a preliminary investigation of this method.

Out of the family of energy functionals described in Section 2, we chose the following one to minimize:
\begin{equation}\label{Edef}
E \equiv E_{(\eta-1)^2} = \int_X\hat\mu\,(\eta-1)^2\,.
\end{equation}
(We remind the reader that here $\hat\mu \equiv (-i)^n\Omega\wedge\bar\Omega$.) %We also evaluated (and performed limited experiments minimizing) the two-derivative functional
%\begin{equation}\label{Ehatdef}
%\hat E[J] = 
%\int_X\hat\mu\,g^{\jb i}\partial_i\ln\eta\,\partial_\jb\ln\eta\,.
%\end{equation}
%However, in general the ultralocal functionals are far easier to evaluate numerically. 
An ultralocal functional was chosen rather than the two-derivative functional $\hat E$ only because it is simpler to evaluate numerically. %(However, as a check on the method, we did evaluate $\hat E$ on the metrics that minimized $E$.) 
Within the family of ultralocal functionals $E_F$ we chose $F = (\eta-1)^2$ because it is smooth, it is simple and therefore easy to evaluate, and it allows us to use the specialized Levenberg-Marquardt algorithm for minimizing sums of squares. Note that, for metrics that are sufficiently close to being Ricci-flat, any generic smooth $F$ is approximately equivalent to $(\eta-1)^2$, since the distribution of $\eta$ over $X$ is highly peaked around 1, and the first relevant term in the Taylor series expansion of $F(\eta)$ about $\eta=1$ is $(\eta-1)^2$ (the constant and linear terms both giving rise to constants when integrated over $X$).

We will refer to the algebraic metrics that minimize $E$ as ``optimal". From \eqref{varEF} and \eqref{amdef} we see that they satisfy
\begin{equation}
\int_X\hat\mu\,\eta\nabla^2\eta\frac{Z^A\bar Z^{\bar B}}{Z^Ch_{C\bar D}\bar Z^{\bar D}} = 0\,.
\end{equation}
For $\eta$ close to 1 this is essentially the condition that the Ricci scalar, integrated against a set of basis functions, vanishes, in other words it can be thought of as a Galerkin condition on the Einstein equation in the form $R=0$. It can be contrasted with the equation defining Donaldson's ``refined" metrics,
\begin{equation}
\int_X\hat\mu\,(\eta-1)\frac{Z^A\bar Z^{\bar B}}{Z^Ch_{C\bar D}\bar Z^{\bar D}} = 0\,,
\end{equation}
which is a Galerkin condition instead on the equation $\eta-1=0$.

As an independent check on the quality of the optimal metrics, we also evaluated the two-derivative functional
\begin{equation}\label{Ehatdef2}
\hat E[J] = 
\int_X\hat\mu\,g^{\jb i}\partial_i\ln\eta\,\partial_\jb\ln\eta = -\frac12\int_X\hat\mu\,R
\end{equation}
on them. (We also experimented in a limited way with minimizing $\hat E$ itself.)

Some specifics of our method for computing the optimal metrics are described in Section 6. Although presently implemented specifically for the CYs described in the next section, it would be straightforward to generalize the implementation to any projectively embedded CY.

%The methods used to obtain the results described in the section are explained in the next one.

\section{Results}

We tested our method on the Fermat quartic
\begin{equation}
\sum_{a=1}^4(z^a)^4 = 0\
\end{equation}
in $\CP^3$, as well as on the quintic
\begin{equation}\label{quintic}
\sum_{a=1}^5(z^a)^5 - 5\psi\prod_{a=1}^5z^a = 0
\end{equation}
in $\CP^4$, for real values of the parameter $\psi$ in the range $-10\le\psi\le10$, which includes the Fermat quintic (at $\psi=0$), and the conifold (at $\psi=1$). These particular CYs were chosen mainly for their simplicity and their high degree of symmetry, which vastly reduces the dimensionality of the parameter space, as we will discuss below. Also, this one-parameter family of quintics includes both singular and uniform examples, allowing us to study the effect of uniformity on convergence (see the discussion at the end of Section 3). Finally, the quintic with $\psi=0.1$ allowed for a direct comparison with the balanced metrics.

\subsection{Fermat quartic}

\FIGURE{
\includegraphics[width=3.5in]{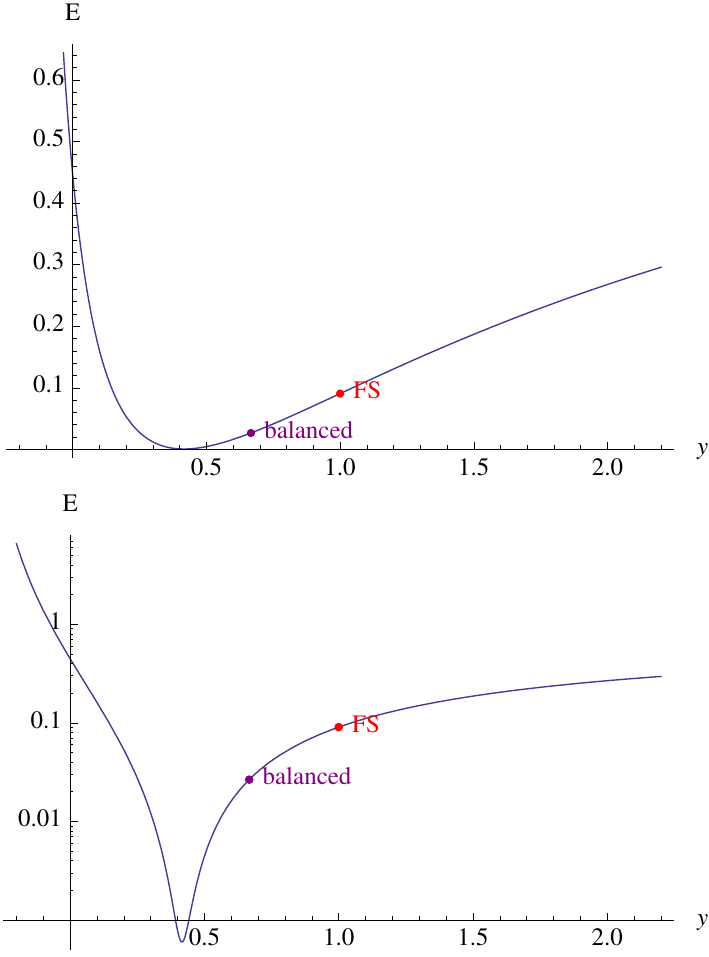}
\caption{$E$ versus $y$ for the $k=2$ algebraic metric on the Fermat quartic, defined by the polynomial \eqref{k2p}, with $E$ plotted on a normal (top) and logarithmic (bottom) scale. For comparison the FS metric (at $y=1$ and $E\approx0.09$) and the balanced metric (at $(y,E)\approx(0.67,0.03)$; see subsection 5.2) are also marked. The minimum (optimal metric) is at $(y,E)\approx(0.42,6\times10^{-4})$.}
\label{Evsyk2}
}

\FIGURE{
\includegraphics[width=4.3in]{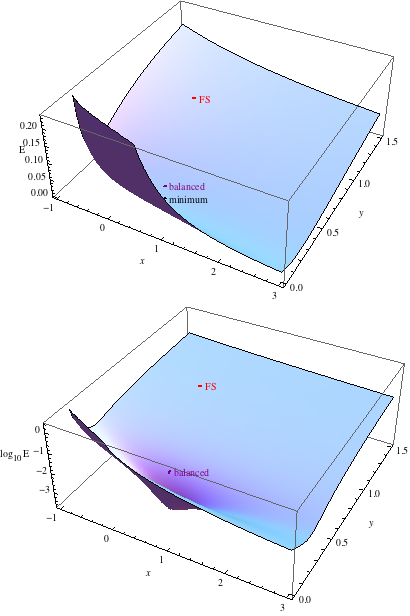}
\caption{$E$ versus $(x,y)$ for the $k=3$ algebraic metric on the Fermat quartic, defined by the polynomial \eqref{k3p}, with $E$ plotted on a normal (top) and logarithmic (bottom) scale. For comparison the FS metric (at $(x,y) = (0,1)$ and $E\approx 0.09$) and balanced metric (at $(x,y,E)\approx (0.43,0.36,0.005)$) are also marked. The minimum (optimal metric) is at $(x,y,E)\approx (0.57,0.27,2\times 10^{-4})$.}
\label{Evsxyk3}
}

We will write the homogeneous degree $(k,k)$ polynomial $Z^Ah_{A\bar B}\bar Z^{\bar B}$ appearing in the \K potential as $p$ (so $K_{(b)}^h=\ln(p^{1/k}/|z^b|^2)$). We know that the Ricci-flat metric is invariant under the full symmetry group of the Fermat quartic, so we may as well restrict attention to algebraic metrics with the same symmetry.\footnote{The symmetry group is of order 3072, and is generated by permutations of the $z^a$, multiplying each by a fourth root of unity (modulo multiplying all by the same root), and complex conjugating all of them. For the Fermat quintic, the symmetry group is analogous, and has order 150,000. For the quintic \eqref{quintic} with $\psi$ real but non-zero, the symmetry group is reduced: multiplying each $z^a$ by a fifth root of unity is only allowed if the product of the roots is 1; the order of this group is 30,000.}  For $k=1$ the only invariant polynomial is the Fubini-Study one, $p=\sum_a|z^a|^2$; for $k=2$ there is a one-parameter family,
\begin{equation}\label{k2p}
p =
y\left(\sum_a\left|z^a\right|^2\right)^2 +
(1-y)\sum_a\left|z^a\right|^4\,;
\end{equation}
while for $k=3$ there are two parameters:
\begin{equation}\label{k3p}
p = 
y\left(\sum_a\left|z^a\right|^2\right)^3 +
x\left(\sum_a\left|z^a\right|^2\right)\sum_a\left|z^a\right|^4 +
(1-x-y)\sum_a\left|z^a\right|^6\,.
\end{equation}
Setting $y=1$ in the $k=2$ case, or $(x,y)=(0,1)$ in the $k=3$ case, returns us to the FS metric. Figure \ref{Evsyk2} shows $E$ versus $y$ for $k=2$, while figure \ref{Evsxyk3} shows $E$ versus $(x,y)$ for $k=3$. The FS metric is marked on these plots for comparison (along with the respective balanced metrics; see subsection 5.2). The purpose of these plots is to show that $E$ is a well-behaved function, that should not present any difficulties for numerical minimization. In addition to being free of local minima,\footnote{Although we showed in subsection 2.1 that $E$ does not have any local minima on the full space of metrics in a given \K class, we have not proven that such minima could not appear when the functional is restriced to the space of algebraic metrics.} it is well-defined and smooth for $y\le0$, despite the fact that the corresponding metrics are actually singular. (As long as there are no local minima, this is an advantage since the minimization algorithm may temporarily jump across to $y\le0$.) The minimum value of $E$ for $k=2$ is about 150 times smaller than the value on the FS metric, corresponding to a more than ten-fold decrease in the spread of $\eta$; going from $k=2$ to $k=3$, $E$ decreases by another factor of about 3. The minimum for $k=2$ occurs at
\begin{equation}
y\approx0.42\,,
\end{equation}
and for $k=3$ at
\begin{equation}
(x,y)\approx(0.57,0.27)\,.
\end{equation}
This $k=3$ optimal metric is probably already a useful rough approximation to the Ricci-flat metric for some practical applications, while being computationally extremely simple.

\FIGURE{
\includegraphics[width=3.5in]{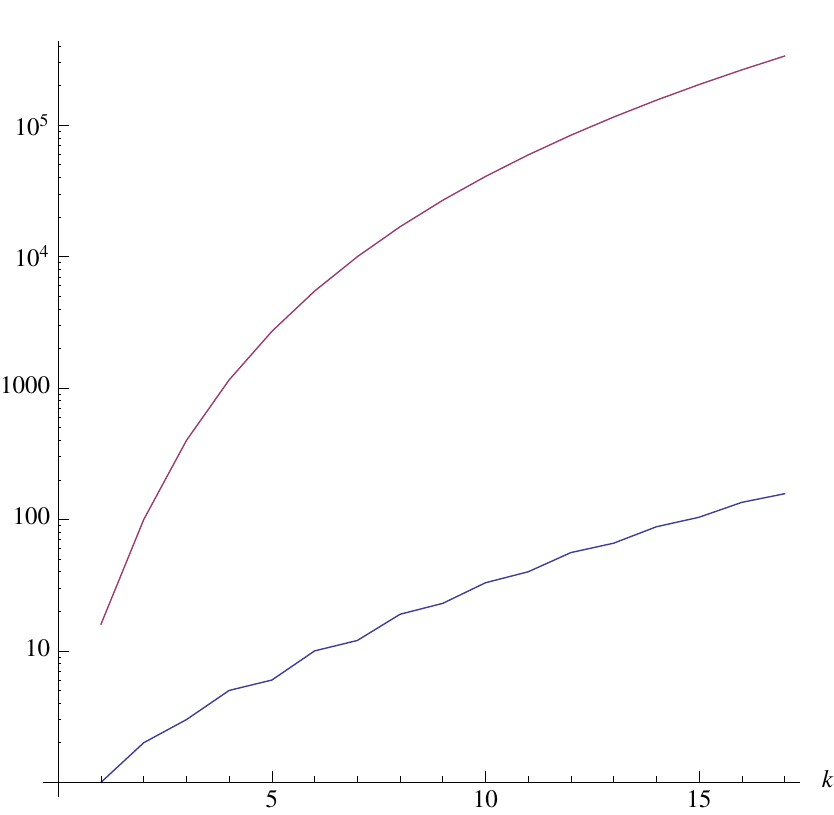}
\caption{Number of coefficients in the degree $(k,k)$ homogeneous polynomial appearing in the algebraic metric, versus $k$, for the quartic. This is the number of real degrees of freedom in the algebraic metric (plus 1, since multiplying $p$ by an overall constant does not change the metric). The top curve is without imposing any symmetry constraints; this is $N_k^{\prime2}=4(k^2+1)^2$ (see equation \eqref{Nkprime}). The bottom curve is after imposing the symmetries of the Fermat quartic (we do not know an explicit formula for this number).}
\label{quarticMvals}
}

\FIGURE{
\includegraphics[width=3.5in]{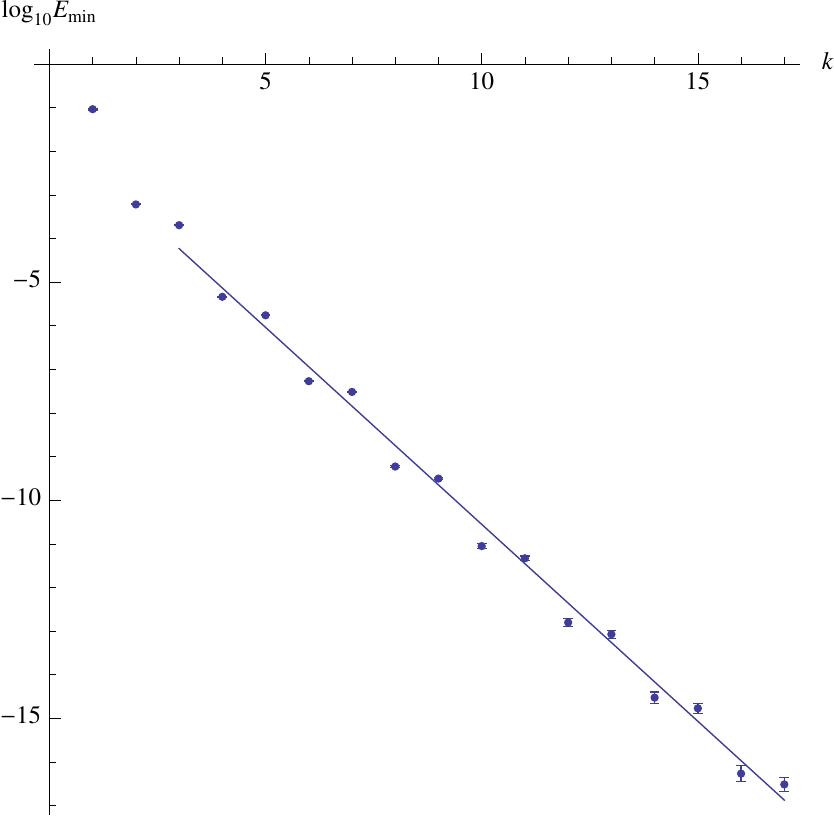}
\caption{Value of $E$ for the optimal algebraic metric on the Fermat quartic, versus $k$. The exponential decrease is clear, accompanied by a small even-odd modulation. The line is $0.03\times 8^{-k}$, obtained by fitting to the points with $k=3$ through $17$.}
\label{quarticEmin}
}

\FIGURE{
\includegraphics[width=6in]{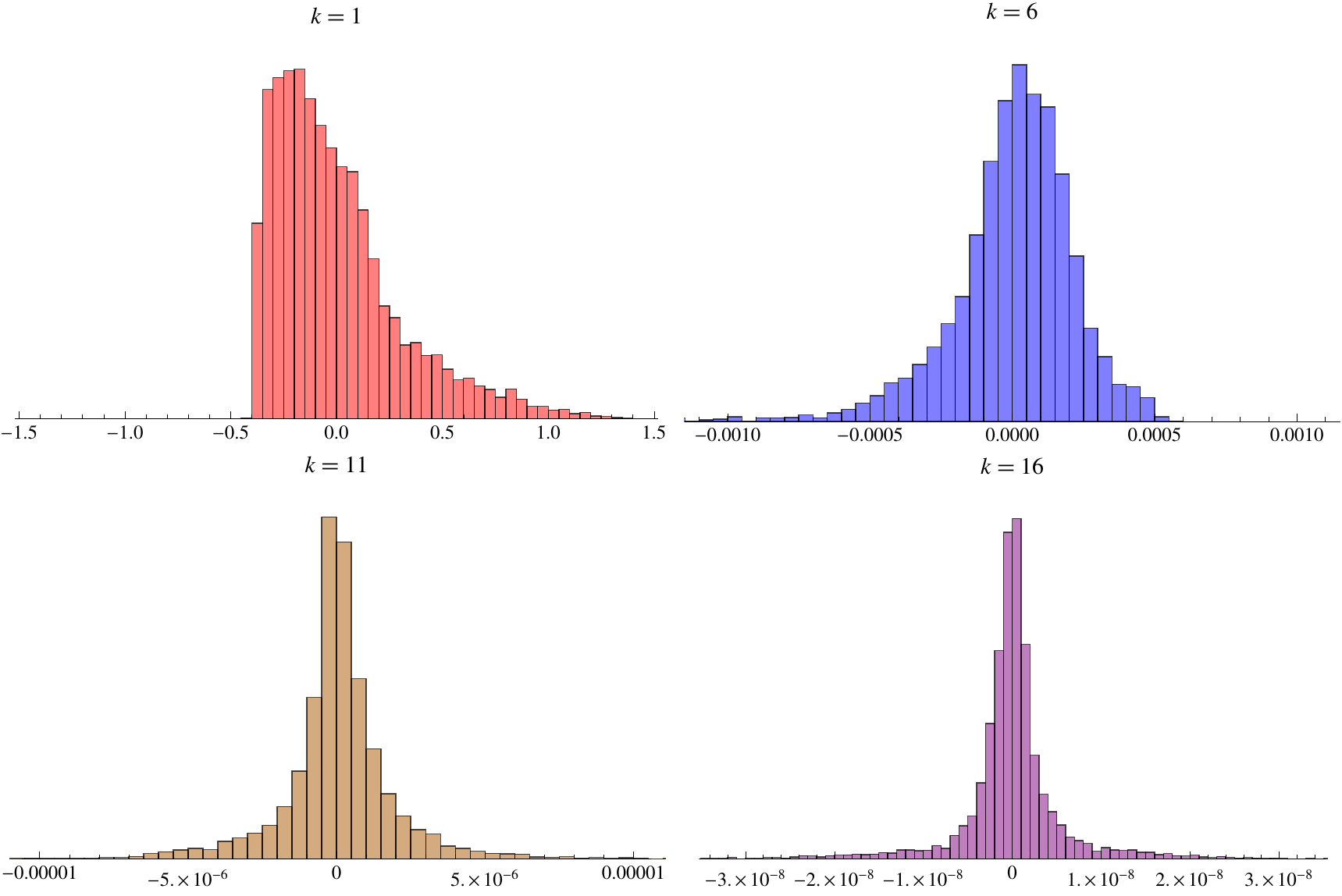}
\caption{Distribution of $\eta-1$ for a set of 9000 points on the Fermat quartic (randomly chosen according to the measure $\hat\mu$), for the optimal metrics with $k=1$, 6, 11, and 16.}
\label{etahistograms}
}

The number of coefficients in $p$ is plotted in figure \ref{quarticMvals}, both with and without imposing imposing invariance under the Fermat symmetries. The value of $E$ for the optimal metric is plotted against $k$ in figure \ref{quarticEmin} for $k$ ranging from 1 to 17. The exponential decrease for large $k$, expected on theoretical grounds (see Section 3), is unmistakeable. Increasing $k$ by 1 leads to a decrease in $E_{\rm min}$ by a factor of about $8$. Superimposed on this exponential decrease is an interesting even-odd modulation, presumably related to the even-odd modulation in the number of parameters evident in figure \ref{quarticMvals}. In figure \ref{etahistograms} we show the distribution of vales of $\eta$ over $X$ for the optimal metrics with $k=1$, 6, 11, and 16.

\FIGURE{
\includegraphics[width=3.5in]{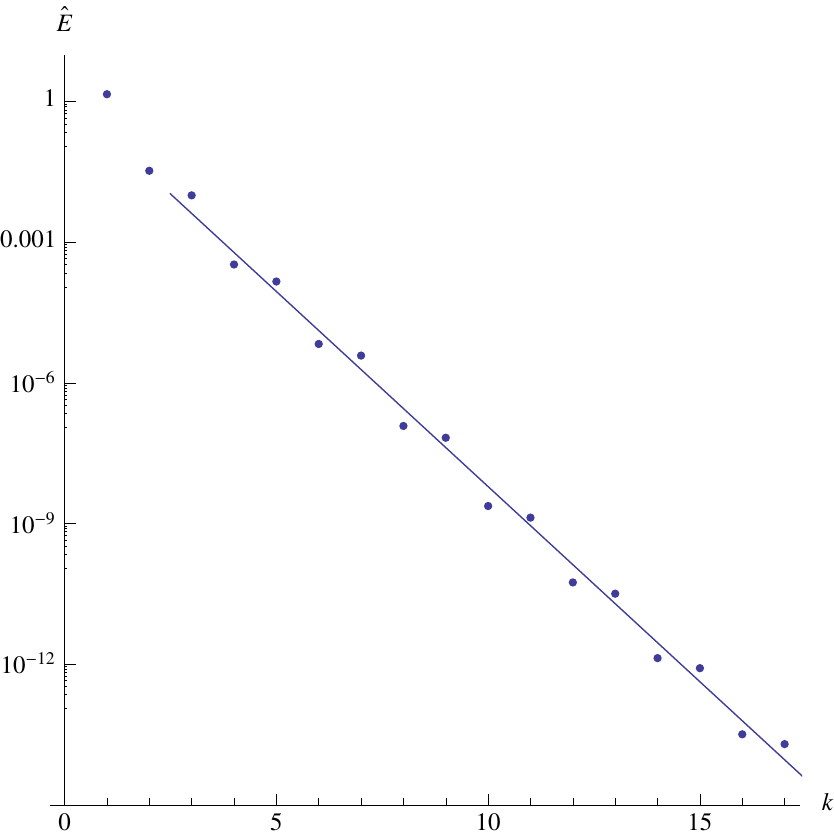}
\caption{Two-derivative functional $\hat E$ \eqref{Ehatdef2} evaluated on the optimal metrics on the Fermat quartic for $k=1$ to 17. Again the exponential decrease is clear. The line is $1.3\times6.8^{-k}$, obtained by fitting to the points with $k=3$ to 17. (The errors are negligible on this scale.)}
\label{quarticEhat}
}

\FIGURE{
\includegraphics[width=6in]{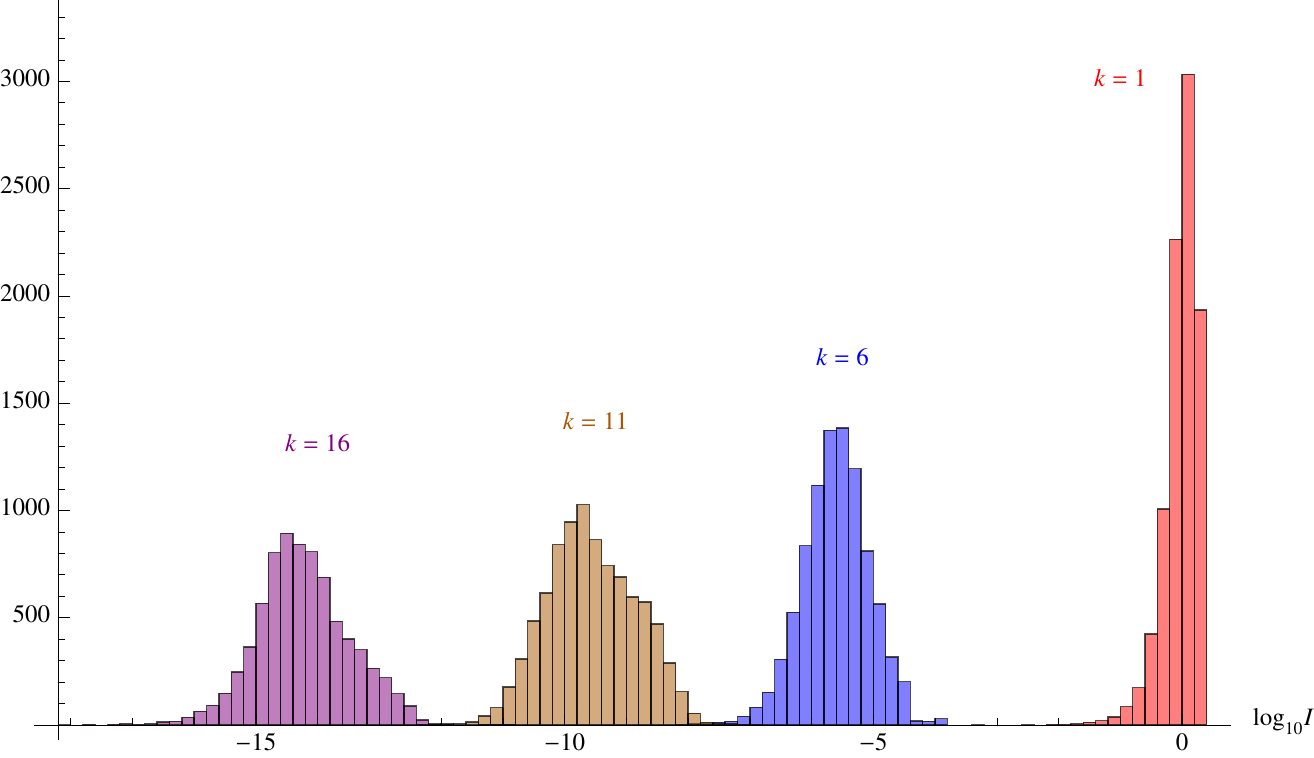}
\caption{Distribution of $I\equiv g^{\jb i}\partial_i\ln\eta\,\partial_\jb\ln\eta$, the integrand of $\hat E$, on a logarithmic scale, for the same 9000 points on the Fermat quartic and the same metrics as in figure \ref{etahistograms}.}
\label{Ihistograms}
}

It is useful to have some independent estimate of the quality of the optimal metrics, other than the value of the functional $E$ that by definition they minimize. As explained below equation \eqref{Edef}, any generic ultralocal functional $E_F$ is essentially equivalent to $E$ for metrics sufficiently close to the Ricci-flat one, so these do not serve the purpose. On the other hand, the functional $\hat E$ \eqref{Ehatdef2} is independent, since it depends on derivatives of the metric; at the same time it is computationally simpler to evaluate than, for example, the Calabi energy $\int \mu_J\,R^2$. $\hat E$ is plotted against $k$ for the optimal metrics in figure \ref{quarticEhat}. Again the exponential decrease is clear, with $\hat E$ decreasing by a factor of about 7 for each unit increase in $k$. Figure \ref{Ihistograms} shows the distribution of values of the integrand of $\hat E$, $I\equiv g^{\jb i}\partial_i\ln\eta\,\partial_\jb\ln\eta$, for $k=1$, 6, 11, and 16.

\FIGURE{
\includegraphics[width=4in]{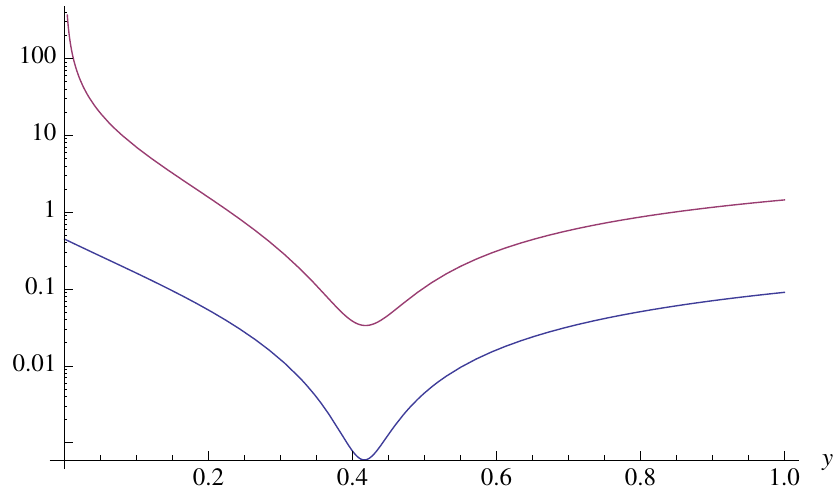}
\caption{$\hat E$ (top curve) and $E$ (bottom curve) versus $y$ for the $k=2$ algebraic metric on the Fermat quartic, defined by the polynomial \eqref{k2p}. Note that, unlike $E$, $\hat E$ goes to infinity at $y=0$, where the metric degenerates. Their minima are separated by less than the error in their respective numerical estimates.}
\label{EEhatk2}
}

Another interesting possibility is to minimize $\hat E$ itself, rather than $E$. This is computationally far more demanding, and we have not carried out a systematic investigation. However, our preliminary experiments indicate that the resulting metrics are not significantly different from those obtained by minimizing $E$, at least for the low values of $k$ we have studied. Figure \ref{EEhatk2} shows both $E$ and $\hat E$ as functions of $y$ for the $k=2$ algebraic metric \eqref{k2p}. The minima of the two functions are clearly closely aligned; in fact they are separated by less than the error on our numerical estimates of them. The same was true for $k=3$ and $4$. For $k=5$ and $6$ (the highest value of $k$ for which we successfully minimized $\hat E$), the minima of $E$ and $\hat E$ were measurably but not significantly separated. For $k=5$, the value of $E$ increased by only about 7\% going from its minimum to that of $\hat E$, and similarly $\hat E$ increased by only about 7\% going from its minimum to that of $E$. For $k=6$, the corresponding increases were even smaller: 3\% and 6\% respectively. However, it is not clear if one can extrapolate to higher values of $k$ (or other manifolds) the conclusion that minimizing $E$ and $\hat E$ are effectively equivalent.

\subsection{Quintics: scanning over moduli space and comparison to balanced metrics}

\FIGURE{
\includegraphics[width=3.5in]{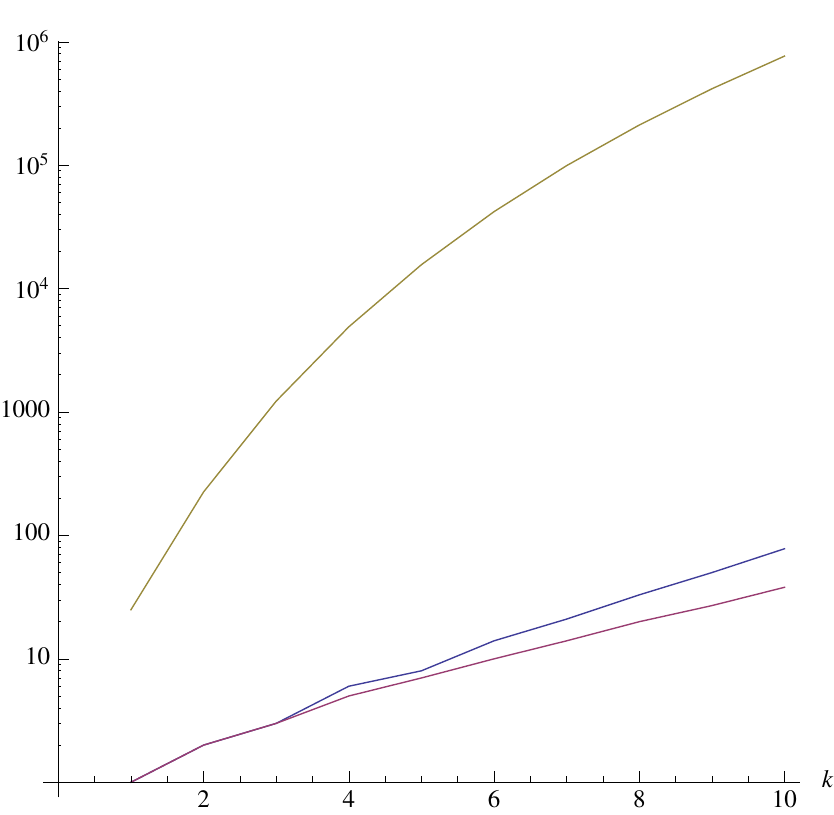}
\caption{Analogue of figure \ref{quarticMvals} for the quintic. The top curve is without imposing any symmetries; the number of coefficients is $N_k^{\prime2} = [\frac56k(k^2+5)]^2$. The bottom curve is after imposing the symmetries of the Fermat quintic. The middle curve is after imposing those of the quintic \eqref{quintic} with $\psi$ real but non-zero.}
\label{quinticMvals}
}

\FIGURE{
\includegraphics[width=4.5in]{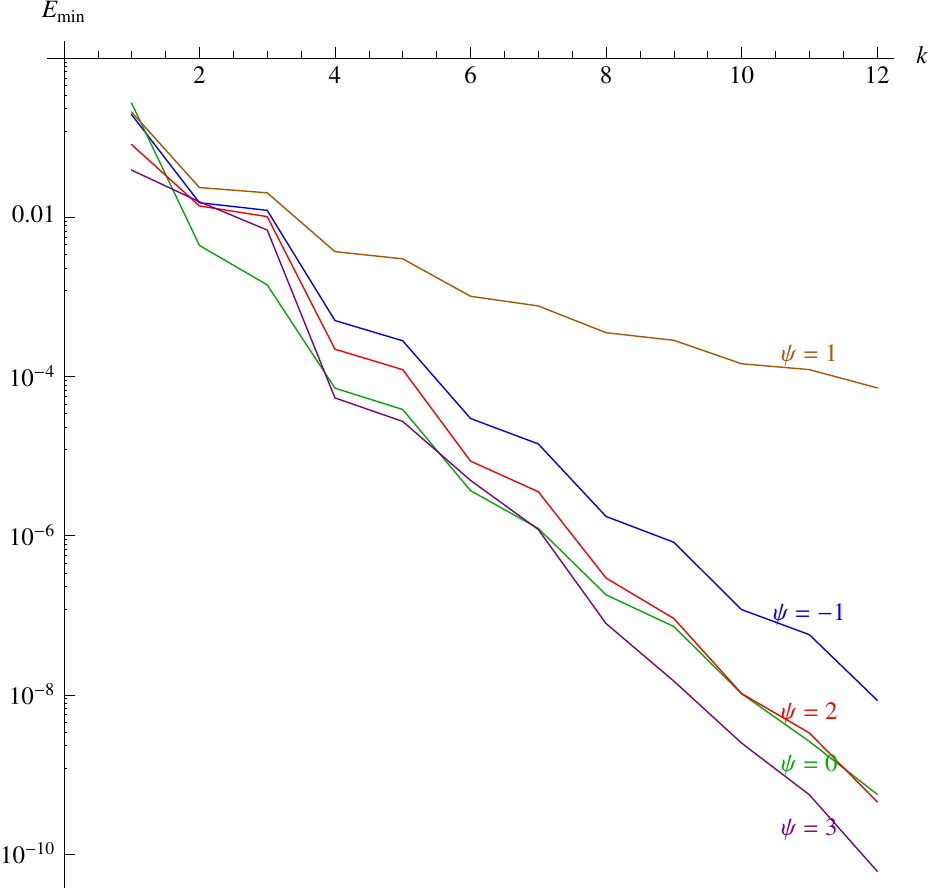}
\caption{$E_{\rm min}$ versus $k$ for algebraic metrics on the quintic \eqref{quintic} for the values of $\psi$ indicated on the plot. The exponential decrease is clear for all values of $\psi$ except $\psi=1$, which is the conifold point in the quintic moduli space. A linear fit of $\log E_{\rm min}$ gives $E_{\rm min}\approx 0.07\times5^{-k}$ (not plotted) for $\psi=0$ (the Fermat quintic).}
\label{quinticEmin}
}

\FIGURE{
\includegraphics[width=4.5in]{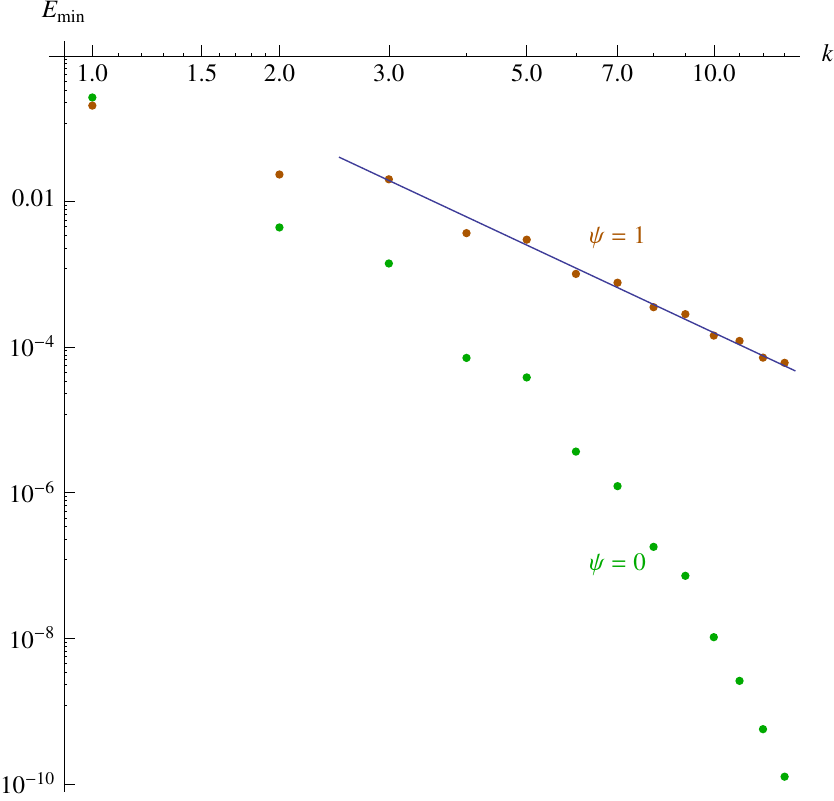}
\caption{Same as figure \ref{quinticEmin}, except only the data for the Fermat and conifold quintics are shown, and $k$ is plotted on a log scale in order to emphasize the power-law behavior of the conifold data. The line is the function $1.6k^{-4}$, obtained by fitting the conifold points to $k^{-4}$. In contrast, the points for the Fermat quintic do not lie on a straight line.}
\label{quinticEminloglog}
}

The analogues of figures \ref{quarticMvals} and \ref{quarticEmin} for the quintic \eqref{quintic}, with $\psi=-1,0,1,2,3$ are shown in figures \ref{quinticMvals} and \ref{quinticEmin}. For all cases except $\psi=1$ (the conifold point), the exponential decrease with $k$ is again clear, with a slightly smaller slope (in absolute value) than for the quartic. (The even-odd modulation seen in the quartic is also still present.) For the conifold quintic, however, the decrease is clearly slower than exponential. Indeed, on a log-log plot (figure \ref{quinticEminloglog}), the conifold points lie approximately on a straight line, with slope $-4$, while the Fermat points lie on a curve. In principle it should be possible to predict the $k^{-4}$ convergence rate from the local geometric structure of the conifold singularity.\footnote{This turns out to be the same as the rate of convergence of the balanced metrics on a smooth CY; see \eqref{balancedsigma} (note that $E\sim\sigma^2$).} This is a concrete example of the discussion of convergence rates in subsection 3.2. These metrics are the first high-accuracy Ricci-flat metrics on the phenomenologically interesting \cite{Baumann:2010sx} compact conifold and deformed conifold.

\FIGURE{
\includegraphics[width=6in]{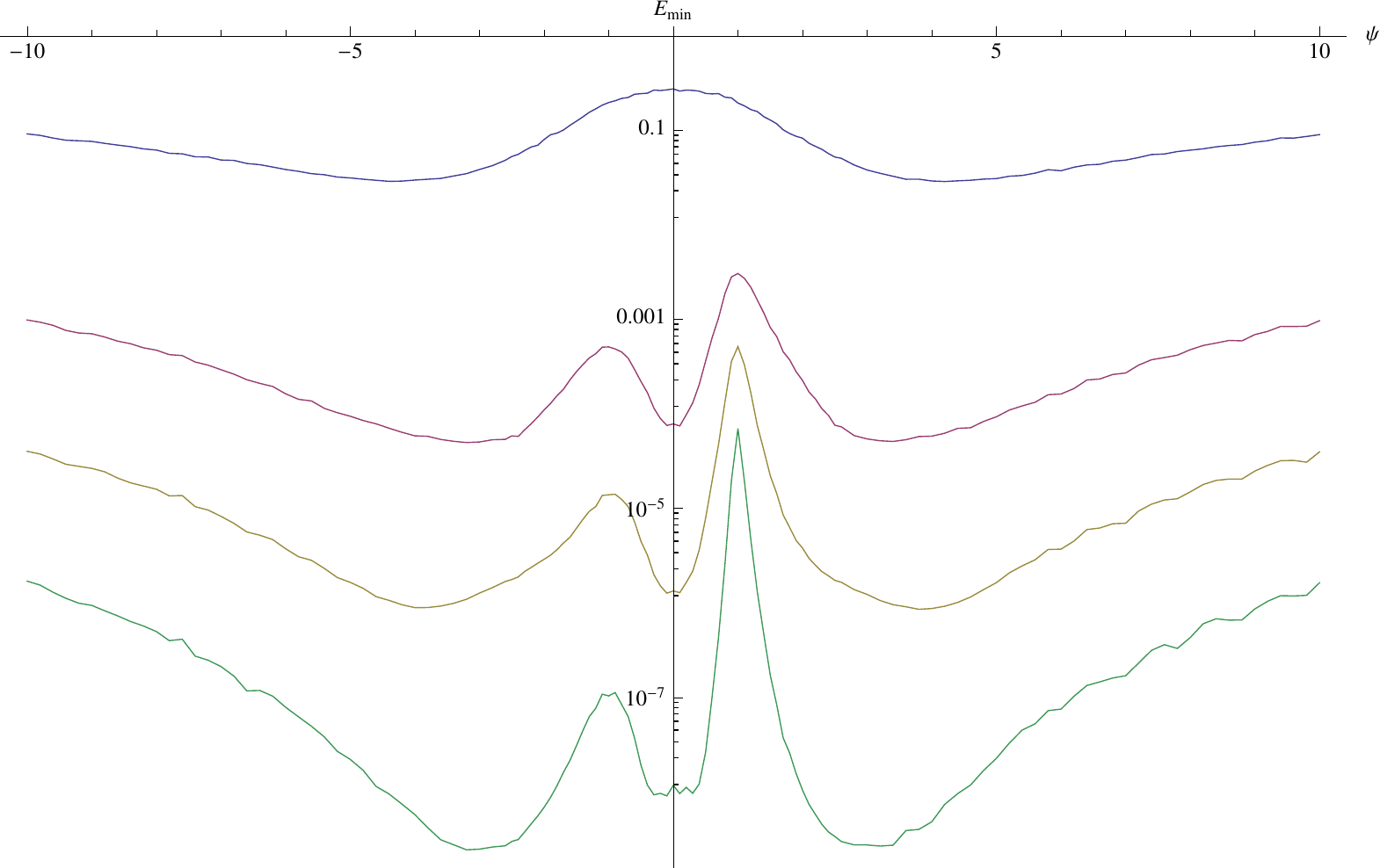}
\caption{$E_{\rm min}$ versus $\psi$ for (top to bottom) $k=1,4,7,10$ (the top curve represents the FS metric). (Error bars are not indicated to avoid cluttering the plot; however, the roughness of the curves, which is due to numerical errors, gives a good indication of their size.)}
\label{psi}
}

Figure \ref{psi} shows how $E_{\rm min}$ varies with $\psi$, for four different values of $k$, in the range $-10\le\psi\le10$. The conifold point stands out in these plots, with a peak whose height and sharpness grow with $k$. (In particular, although the induced FS metric, $k=1$, is singular, that singularity does not show up in $E$, reflecting the ultralocality of that functional---the induced FS metric has a curvature singularity but an everywhere finite ratio $\eta$ of volume forms.) There is also a smaller and less sharp bump at $\psi =-1$, presumably due to the proximity in the complex $\psi$ plane of the points $\psi = e^{\pm4\pi i/5}$, which also represent the conifold point. (Recall that, as a coordinate on the moduli space, $\psi\sim e^{2\pi i/5}\psi$.) It is also interesting to note that, while the Fermat quintic is a local minimium of $E_{\rm min}$ with respect to $\psi$, the global minimum is actually past the conifold point, at $\psi\approx3$. However, for large (positive and negative) values of $\psi$, $E_{\rm min}$ grows rapidly, as one expects from the fact that the quintic is singular in the large complex structure limit.

\FIGURE{
\includegraphics[width=3.5in]{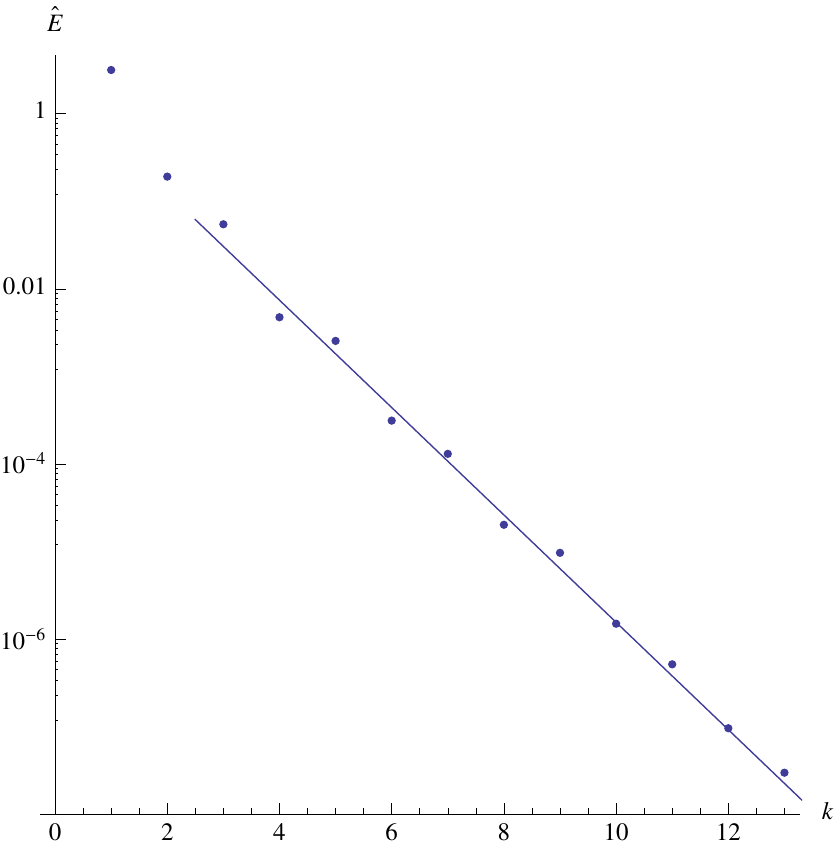}
\caption{Two-derivative functional $\hat E$ \eqref{Ehatdef2} evaluated on the optimal metrics on the Fermat quintic for $k=1$ to 13. The line is $2.1\times4.1^{-k}$, obtained by fitting to the points with $k=3$ to 13.}
\label{quinticEhat}
}

\FIGURE{
\includegraphics[width=5.5in]{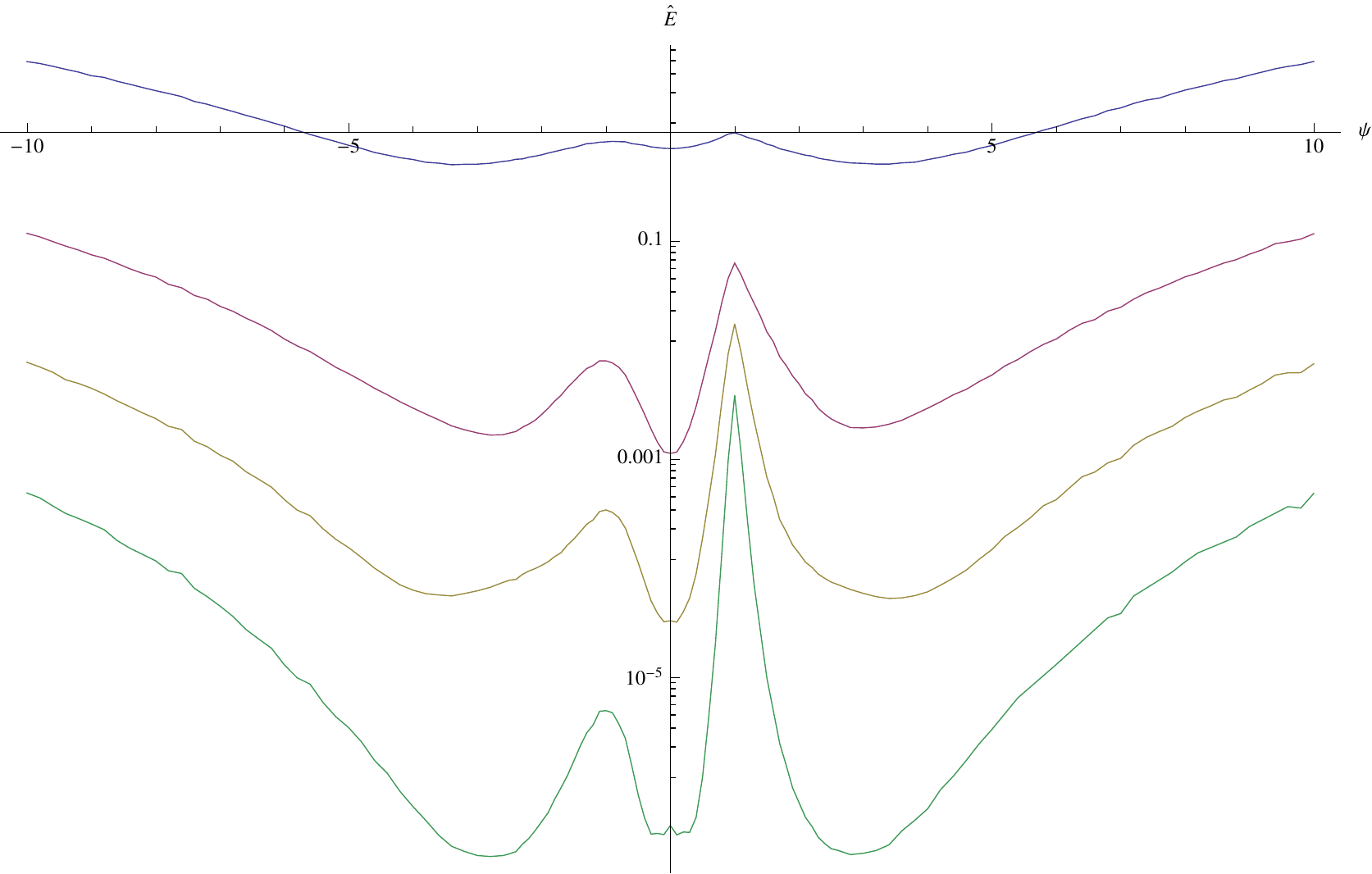}
\caption{Same as figure \ref{psi}, except $\hat E$ instead of $E_{\rm min}$.}
\label{quinticEhatvspsi}
}

Figure \ref{quinticEhat} is the analogue of figure \ref{quarticEhat} for the Fermat quintic ($\psi=0$). Again the exponential decrease is clear, with a somewhat smaller slope. And figure \ref{quinticEhatvspsi} shows $\hat E$ plotted against $\psi$ for the same values of $\psi$ and $k$ as figure \ref{psi}. It shows qualitatively similar features to that figure. Perhaps the most interesting difference is in the $k=1$ curve, representing the FS metric; unlike $E$, the two-derivative functional is sensitive to the fact that the induced FS metric is singular at the conifold point and at $\psi=\infty$.

\FIGURE{
\includegraphics[width=3.5in]{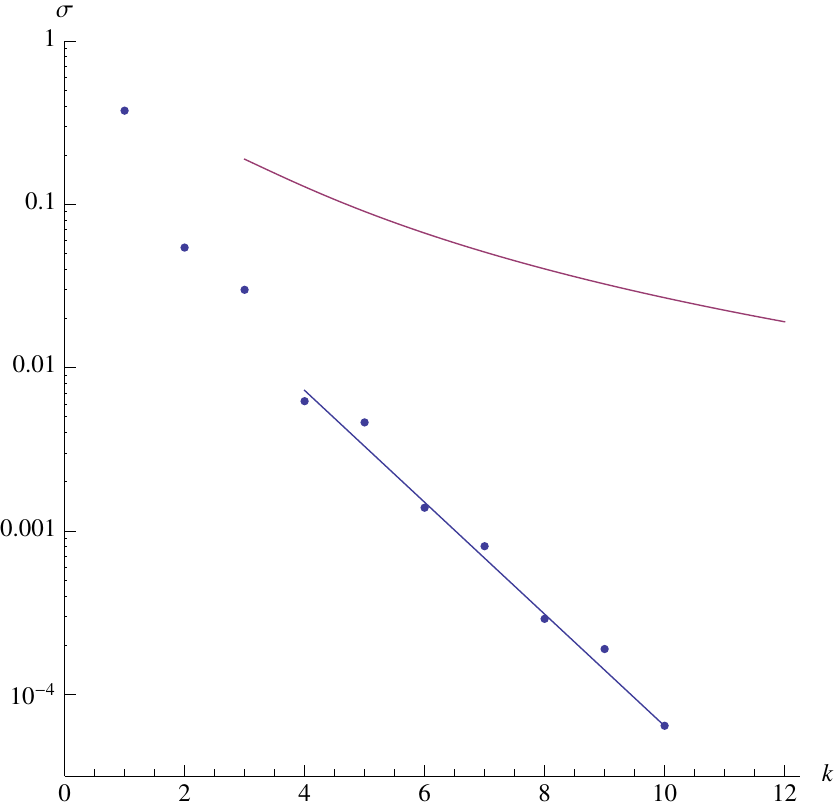}
\caption{Error functional $\sigma$ defined in \eqref{sigmadef} versus $k$. Blue points: optimal metrics (the errors are negligible on this scale). Blue line: fit \eqref{optimalsigma} to those points for $k=4$ through $10$. Purple curve: fit \eqref{balancedsigma} reported in \cite{Douglas:2006rr} for the balanced metrics for $k=3$ through $12$.}
\label{balanced}
}

Finally, it is interesting to compare these optimal metrics to the balanced metrics \cite{MR2508897}. For comparison to the optimal metrics, the balanced metrics on the Fermat quartic for $k=2$ and 3 are marked on figures \ref{Evsyk2} and \ref{Evsxyk3} respectively. The balanced metrics were thoroughly studied as approximations to the Ricci-flat metric in a series of papers by Douglas and collaborators (along with generalizations to the hermitian Yang-Mills equation and related physical issues) \cite{Douglas:2006hz,Douglas:2006rr,Braun:2007sn,Braun:2008jp}. In particular, Douglas, Karp, Lukic, and Reinbacher published enough details to allow a direct comparison for the quintic \eqref{quintic} with $\psi=0.1$ \cite{Douglas:2006rr}. They reported that the following error functional,\footnote{The functional $\sigma$ has the same form as the ultralocal functionals $E_F$. However, since $|\eta-1|$ is neither a smooth nor a strictly convex function, it is probably not an ideal choice for numerical minimization.}
\begin{equation}\label{sigmadef}
\sigma[J] \equiv \int\hat\mu\,|\eta-1|\,,
\end{equation}
evaluated on the balanced metrics with $k=3$ through $12$, was given approximately by
\begin{equation}\label{balancedsigma}
\sigma \approx 3.1k^{-2} - 4.2k^{-3}\qquad \hbox{(balanced)}.
\end{equation}
In figure \ref{balanced} we plot this function, together with the value of $\sigma$ for the optimal algebraic metric for $k=1$ through 10; the latter values follow, as discussed above, an exponential decrease, with the following fit:
\begin{equation}\label{optimalsigma}
\sigma \approx 0.17\times 2.2^{-k} \qquad \hbox{(optimal)}.
\end{equation}
From this plot it is clear that the optimal metrics are a significant improvement over the balanced ones, whether one is interested in a rough-and-ready approximation to the Ricci-flat metric (low $k$) or a high-precision one (high $k$). Although we have not carried out a thorough comparison of the two methods from the computational viewpoint (i.e.\ ease of implementation and computational resources required), our experience, together with that reported by Donaldson \cite{MR2508897} and Douglas and collaborators \cite{Douglas:2006hz,Douglas:2006rr,Braun:2007sn,Braun:2008jp} suggests that the optimal metrics are no harder and perhaps somewhat easier to compute than the balanced metrics.

\section{Methodology}

The problem of finding the optimal metrics can be divided into three parts:
\begin{itemize}
\item Constructing the relevant space of polynomials $p$ appearing in the \K potential for a given $k$.
\item Evaluating the functional $E$ for a given $p$.\footnote{While we will not describe it explicitly here, the computation of the two-derivative functional $\hat E$ \eqref{Ehatdef} can be done following the same strategy as for $E$. In practice this computation is significantly more complicated. In particular, analytic computation of the gradient of $\hat E$ is prohibitively complicated, rendering its numerical minimization significantly more difficult.}
\item Minimizing $E$ over the space of $p$'s.
\end{itemize}
We will explain our approach to each part in turn. All computations were performed in \emph{Mathematica}, with a code of a few hundred lines,\footnote{This code is available for download at \href{http://people.brandeis.edu/~headrick/physics/}{\tt http://people.brandeis.edu/~headrick/physics/}, along with a notebook illustrating its use.} running on an ordinary laptop.

Some aspects of the methods described below are similar to ones employed previously in numerical studies of algebraic metrics \cite{MR2508897,Douglas:2006hz,Douglas:2006rr,Braun:2007sn,Braun:2008jp}.

\subsection{Constructing the space of polynomials}

Our first task is to construct a basis for the relevant space of polynomials $p$ that enter into the algebraic metric. This is the space of real-valued homogeneous degree $(k,k)$ polynomials in the $z^a$ that are invariant under the symmetry group of $X$, modulo those that vanish identically on it. Let $P$ be the defining polynomial of $X$ and $\Gamma$ its symmetry group. For the CYs studied in this paper, $P$ had only real coefficients, and therefore $\Gamma$ included as a subgroup a $\Z_2$ group generated by complex conjugation of the $z^a$. (The method described here easily generalizes to the situation where this is not the case.) Let $\tilde\Gamma$ be the subgroup of $\Gamma$ that acts holomorphically on the $z^a$ (in our case, by permuting the $z^a$ and multiplying them by $(N+1)$ roots of unity). The real-valued $\Gamma$-invariant $(k,k)$ polynomials form a real vector space $\Pi_k$. If $k>N$, then this space contains a non-trivial subspace $\Pi^0_k$ of polynomials that vanish identically on $X$; such polynomials are of the form $gP+\bar g\bar P$, where $g$ is a degree $(k-N-1,k)$ $\tilde\Gamma$-invariant polynomial. We wish to construct a basis for $\Pi_k/\Pi^0_k$.

This problem may be solved using techniques from commutative algebra, namely by constructing the primary and secondary invariants of $\Gamma$ in the polynomial ring $\C[z^a,\bar z^a]$. (An example of such a construction is given in \cite{Braun:2007sn}.) In practice, however, we found it easier to follow a less mathematically sophisticated but more direct route, and have \emph{Mathematica} construct the necessary basis separately for each $k$, by: (1) constructing bases for $\Pi_k$ and $\Pi^0_k$ in terms of symmetrized monomials; (2) expressing the basis elements of $\Pi^0_k$ in terms of those of $\Pi_k$; (3) using the built-in function {\tt NullSpace} to construct a basis for $\Pi_k/\Pi^0_k$. The process takes a few seconds.

\subsection{Evaluating the energy functional}

We now turn to the evaluation, for a given $p$, of the functional
\begin{equation}
E = \int_X\hat\mu\,(\eta-1)^2\,.
\end{equation}
This task can be subdivided into the problem of evaluating $\eta$ at a given point, and the problem of integrating over $X$.

First, however, let us address the small matter of the normalizations. As noted in footnote \ref{normalization}, the algebraic metric is in the same class as the FS metric for any choice of $k$ and $p$, but this class is not volume-normalized. Indeed, we do not have an analytic formula for the volume of $X$ in this class, but must evaluate it numerically. Similarly, defining $\hat\mu$ using the standard formula for $\Omega$ for a projectively embedded CY, we do not know $\int_X\hat\mu$ analytically. In practice, therefore, rather than calculate $\eta$ itself we calculate the unnormalized volume ratio
\begin{equation}
v \equiv \frac{\mu_J}{\hat\mu}\,.
\end{equation}
Defining the average with respect to $\hat\mu$,
\begin{equation}
\langle f\rangle_{\hat\mu} \equiv \frac{\int_X\hat\mu\,f}{\int_X\hat\mu}
\end{equation}
(so that $\eta = v/\langle v\rangle_{\hat\mu}$), we then calculate $E$ as the normalized standard deviation of $v$:
\begin{equation}\label{newEdef}
E = \frac{\left\langle\left(v-{\langle v\rangle_{\hat\mu}}\right)^2\right\rangle_{\hat\mu}}{\langle v\rangle_{\hat\mu}^2}\,.
\end{equation}

Thanks to the simplicity of the algebraic metrics, the evaluation of $v$ may be done ``exactly" (i.e.\ up to round-off errors): no finite-differencing or other numerical approximations are necessary to evaluate the derivatives involved. Using the standard formula for $\Omega$ for a CY embedded in \CPN, we can derive the following useful formula for $v$. For notational concreteness let us assume that we are in the patch $O_{(N+1)}$ of \CPN, i.e.\ where $z^{N+1}\neq0$, and set $z^{N+1}=1$. The defining polynomial $P$ and the mixed polynomial $p$ are then functions of the $N$ coordinates $z^\alpha$, where $\alpha = 1,\ldots,N$ (and, in the case of $p$, their conjugates). We define the $(N+1)\times(N+1)$ matrix $\Psi_{a\bar b}$ by
\begin{equation}
\Psi_{\alpha\bar\beta} = 
\frac{\partial^2p}{\partial z^\alpha\partial\bar z^{\bar\beta}}\,, \qquad
\Psi_{\alpha,N+1} = 
\frac{\partial p}{\partial z^\alpha}\,, \qquad
\Psi_{N+1,\bar\beta} = 
\frac{\partial p}{\partial\bar z^{\bar\beta}}\,, \qquad
\Psi_{N+1,N+1} = p\,,
\end{equation}
and the $(N+1)$-vector $Q_a$ by
\begin{equation}
Q_\alpha = \frac{\partial P}{\partial z^\alpha}\,, \qquad
Q_{N+1} = 0\,.
\end{equation}
We then have
\begin{equation}\label{vformula}
v = k^{1-N}\,p^{-N}\det\Psi\,\bar Q_{\bar b}\Psi^{\bar ba}Q_a\,,
\end{equation}
where $\Psi^{\bar ba}$ is the inverse of $\Psi_{a\bar b}$. The fact that $v$ can be computed essentially to machine precision is very important. Note that the distribution of values of $v$ over $X$ has a spread as small as $10^{-8}$ in some cases (see figure \ref{quarticEmin}; the spread in $v$ is roughly $E^{1/2}$); such precision would be meaningless in the presence of significant numerical errors from, for example, finite-differencing.

Unfortunately, we do not know how to analytically integrate functions over $X$ (even the constant function, let alone $v$, which is a rational function), so the integration must be performed numerically. We have chosen to carry it out by a Monte-Carlo method, for several reasons: it is simple to implement; it allows us to work directly in \CPN, rather than having to construct coordinate patches on $X$; and it has well-understood errors.\footnote{The papers \cite{Douglas:2006hz,Douglas:2006rr,Braun:2007sn,Braun:2008jp} contain discussions of various Monte-Carlo schemes for integration on $X$.} However, we believe that it may be possible to achieve significant improvements in this area; in fact, the integration is the limiting step of the whole method, in terms of accuracy, efficiency, and memory requirements.

The Monte Carlo method instructs us to pick a set of points on $X$ randomly according to the measure $\hat\mu$, and replace the integral with an average over the points, i.e.\ replace $\langle\,\cdot\,\rangle_{\hat\mu}$ with $\langle\,\cdot\,\rangle_{\mu'}$, where $\mu'$ is the sum of point masses. In our implementation, the points on $X$ were generated by a rejection method. We start by dividing \CPN into patches that overlap only on sets of measure zero, for example $O_{(c)}\equiv\{|z^\alpha|^2/|z^c|^2\le1\,\forall\alpha\neq c\}$. As in the previous paragraph, let us for notational concreteness work in the patch $O_{(N+1)}$ and set $z^{N+1}=1$. We have the following formula for $\hat\mu$:
\begin{equation}
\hat\mu = (-i)^n\delta^2(P)\prod_\alpha dz^\alpha\wedge\prod_\beta d\bar z^{\bar\beta}\,.
\end{equation}
We then spread out the delta function a bit:
\begin{equation}
\hat\mu_{\epsilon,\CP^N} = (-i)^n\Theta(\epsilon-|P|)\prod_\alpha dz^\alpha\wedge\prod_\beta d\bar z^{\bar\beta}
\end{equation}
for some small parameter $\epsilon$ ($\Theta$ is the Heaviside function). The measure $\hat\mu_{\epsilon,\CP^N}$ thus has support on a slab of thickness $\epsilon$ about $X$ in \CPN. To choose points according to $\hat\mu_{\epsilon,\CP^N}$, we choose points randomly in $O_{(N+1)}$ according to the coordinate measure, and throw out those that do not satisfy $|P|<\epsilon$. The remaining points are then projected onto $X$ orthogonally with respect to the FS metric. The projected points are distributed according to the measure $\hat\mu_{\epsilon,X}$ obtained by projecting $\hat\mu_{\epsilon,\CP^N}$ onto $X$; they define the measure $\mu'$. This procedure introduces a systematic error of order $\epsilon^2$,
\begin{equation}\label{epserror}
\hat\mu_{\epsilon,X} = \hat\mu + O(\epsilon^2)_{\rm systematic}\,,
\end{equation}
as well as a random error of order $N_{\rm points}^{-1/2}$,
\begin{equation}\label{Nerror}
\mu' = \hat\mu_{\epsilon,X} + O(N_{\rm points}^{-1/2})_{\rm random}\,.
\end{equation}

Now, a crucial fact is that these errors in approximating $\hat\mu$ do \emph{not} lead to comparable errors in the computed metrics. The reason is that they do not enter into the computation of $v$, which is done using the \emph{exact} measure $\hat\mu$. Rather, they enter only in the evaluation of the integral, which becomes $\int_X\mu'(\eta-1)^2$ rather than $\int_X\hat\mu(\eta-1)^2$. The perhaps surprising point is that, in defining our functional $E$, we needn't have used $\hat\mu$; indeed, for \emph{any} fixed volume form $\mu$, the functional $\int_X\mu(\eta-1)^2$ is well-behaved and minimized precisely on the Ricci-flat metric (even more generally, so is any functional of the form $\int_X\mu\,F(\eta)$, so long as $F'(1)=0$). Hence, strictly speaking, the errors in \eqref{epserror}, \eqref{Nerror} aren't errors at all; they just define different, equally good choices of energy functional. The main reason to integrate using $\hat\mu$ is that it is canonical. While our method forces us to use a finite number of points, we do not wish our so-called optimal metric to depend too much on which points happen to be randomly selected. The criterion for choosing $N_{\rm points}$ was thus that different samples of points should lead to metrics that differ by a negligible amount. (The error estimates are discussed quantitatively in subsection 6.4.) In practice we used\footnote{This appears to be substantially smaller than the number of points required to accurately compute the balanced metrics \cite{Douglas:2006hz,Douglas:2006rr,Braun:2007sn,Braun:2008jp}.}
\begin{equation}
N_{\rm points} = 3000
\end{equation}
for all the data shown in the previous section (the larger $k$ is, the more points are required to meet this criterion; a smaller number than 3000 would have sufficed for the lower values of $k$). We also chose\footnote{Except  for the data required to generate figures \ref{psi} and \ref{quinticEhatvspsi}, for which we chose $\epsilon = 0.03$, still small enough to ensure that the $\epsilon$ error is much smaller than the $N_{\rm points}$ error. There is a cost in time to choosing smaller values of $\epsilon$, since only of order $\epsilon^2$ points on \CPN are unrejected. For most of our experiments, the time required to generate the sample of points was small compared to the times required for other parts of the computations. When scanning the moduli space, however, that was not the case, since we required a fresh set of points at each point in moduli space.}
\begin{equation}
\epsilon = 0.01\,,
\end{equation}
ensuring that the $\epsilon$ error was much smaller than the $N_{\rm points}$ error. Generating a sample of 3000 points with this value of $\epsilon$, on either the quartic or the quintic, required on the order of 10 minutes. To reiterate: to claim accuracy in the metric at the level of $10^{-8}$, it is not necessary to evaluate the integrals to this accuracy, which would have required $\epsilon\approx10^{-4}$ and $N_{\rm points}\approx10^{16}$, which is obviously unattainable.

The minimization of $E$ requires the repeated evaluation of $v$ at each point for different choices of $p$. The computation of $v$ requires the matrix $\Psi$ and the vector $Q$. $Q$ is independent of $p$, and can therefore be computed once for each point and stored. Furthermore, $\Psi$ is linear in $p$, and can therefore be computed once, and stored, for each basis polynomial and each point. The time required to compute this data depends on $k$; for a sample of 3000 points and the values of $k$ studied here, it takes from seconds to hours. Storing it requires from around 1 MB to around 100 MB. With this data in hand, evaluating $E$ for a given $p$ takes on the order of a tenth of a second.

Because the integration is performed by evaluating the integrand at a large number of points in real space, the method described here is not a true spectral method, despite the fact that it employs a spectral representation. The amount of data associated with this large number of points (the data described in the previous paragraph, along with other data generated by the function minimization procedure) grows with $k$, and storage is ultimately the limiting factor in going to high values of $k$. (In principle one could compute this data, and even generate samples of points, on the fly, rather than storing it in RAM. However, in practice this is even slower than storing the data on a hard drive and swapping it in and out of RAM.)

\subsection{Minimizing the energy functional}

The numerical minimization of $E$ was carried out using the Levenberg-Marquardt algorithm as implemented in the built-in  \emph{Mathematica} function {\tt FindMinimum}. This algorithm is specially adapted to functions that are sums of squares.\footnote{See, for example, Section 15.5 of \cite{nr} for an explanation of the Levenberg-Marquardt algorithm.} In practice, for simplicity, the function minimized was actually the numerator of \eqref{newEdef}, as the denominator is independent of $p$ up to numerical errors. The algorithm requires not only the value of $v$ at each point but also its derivatives with respect to the parameters being varied (the coefficients in the polynomial $p$). A formula for these derivatives can be derived from \eqref{vformula}, allowing them, like $v$, to be computed essentially to machine precision.

In all cases the minimization was initialized on the FS metric (which is an algebraic metric for any $k$). The time required to perform the minimization ranged from a few seconds for low values of $k$ to around an hour for the highest values.

\subsection{Error estimates}

The error on the minimum value of $E$, shown by the error bars in figure \ref{quarticEmin}, was estimated by a bootstrap method as follows. Three independent samples of 3000 points were generated. $E$ was minimized using each sample, yielding three different metrics. $E$ was then evaluated on each metric using each sample, for a total of nine different values. The (logarithmic) mean and standard deviation of these nine values are respectively the central value and the error for $E_{\rm min}$ reported on the plot.

\FIGURE{
\includegraphics[width=3.42in]{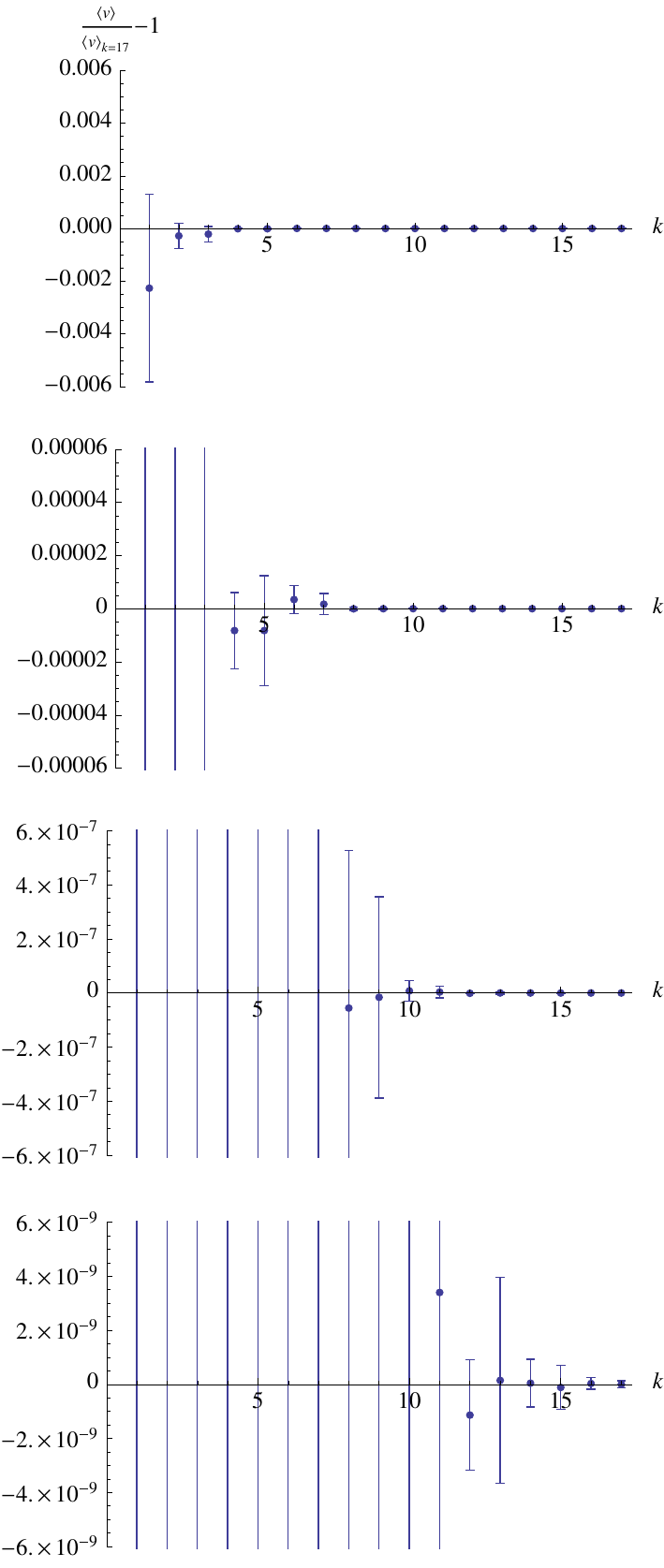}
\caption{Relative deviation of $\langle v\rangle_{\mu'}$ from its value at $k=17$, versus $k$, for the optimal metrics on the Fermat quartic. The only difference between the plots is the scale of the vertical axis. The errors are estimated using a bootstrap, as described in the text. The plots show consistency, within the errors, in the value of $\langle v\rangle_{\mu'}$.}
\label{vconsistency}
}

Finally, we give a consistency check on our calculations, taking advantage of the fact that the distribution of $v$ over $X$ for the optimal metric becomes highly peaked as $k$ increases. By definition, its normalized standard deviation is $E^{1/2}$, which, for the Fermat quartic, is less than $10^{-8}$ for $k=17$ (the highest value for which the computation was done). The theoretical mean value $\langle v\rangle_{\hat\mu}$ is independent of the metric, but the computed value $\langle v\rangle_{\mu'}$ has an error on the order of $E^{1/2}N_{\rm points}^{-1/2}$ (not counting the $\epsilon$-error, which is of order $E^{1/2}\epsilon^2$ and therefore negligible in our computations). Figure \ref{vconsistency} shows the normalized differences between $\langle v\rangle_{\mu'}$ at different $k$ and its value at $k=17$, showing that $\langle v\rangle_{\mu'}$ remains constant within its error, even as that error decreases by more than seven orders of magnitude. The error bars in the figure were computed using the bootstrap method described in the previous paragraph, and agree in order of magnitude with the parametric estimate $E^{1/2}N_{\rm points}^{-1/2}$.

\acknowledgments

We would like to thank V. Braun, M. Douglas, R. Karp, B. Lian, A. Lawrence, W. Nahm, A. Neitzke, G. Schwarz, Y. Tachikawa, A. Tomasiello, and especially T. Wiseman for very helpfulcomments and discussions. A substantial part of this work was done while M.H. was supported by a Pappalardo Fellowship at the MIT Center for Theoretical Physics. M.H. would also like to thank the KITP and the Harvard Center for the Fundamental Laws of Nature for hospitality while this work was being completed. This work was supported in part by DOE grant No.\ DE-FG02-92ER40706.

\bibliography{ref}
\bibliographystyle{JHEP}

\end{document}